\documentclass[aps,pra,reprint,superscriptaddress,letterpaper,twocolumn,longbibliography]{revtex4-1}
\usepackage[english]{babel}
\usepackage{ucs}
\usepackage[utf8x]{inputenc}
\usepackage{amsmath}
\usepackage{mathtools}
\usepackage{amssymb}
\usepackage{amsthm}
\usepackage{mathrsfs}
\usepackage{graphicx}
\usepackage{bm}
\usepackage{bbm}
\usepackage{empheq}
\usepackage{cases}
\usepackage{euscript}
\usepackage[usenames, dvipsnames, x11names]{xcolor}
\usepackage[colorlinks=true,linkcolor=SpringGreen4,citecolor=blue,urlcolor=Magenta3]{hyperref}
\usepackage[shortlabels]{enumitem}
\newcommand{\bra}[1]{\langle #1|}
\newcommand{\ket}[1]{|#1\rangle}

\newcommand{\ketbra}[2]{\ket{#1}\!\bra{#2}}

\newcommand{\mm}[1]{\mathrm{#1}}
\newcommand{\abs}[1]{\left|#1\right|}

\newcommand{\di}[1]{\mathop{}\!\mathrm{d} #1}

\newcommand{\s}[1]{\hat{\sigma}_{#1}}
\newcommand{\n}[1]{\hat{\nu}_{#1}}
\newcommand{\la}[1]{\hat{\lambda}_{#1}}
\newcommand{\simpletable}[2]{ \begin{tabular}{@{}l@{}} #1 \\ #2 \end{tabular} }
\def \ua{\mathrm{a}}

\def \uc{\mathrm{c}}

\def \uq{\mathrm{q}}

\def \uf{\mathrm{f}}
\def \ud{\mathrm{d}}
\def \uq{\mathrm{q}}
\def \uI{\mathrm{I}}
\def \uG{\mathrm{G}}
\def \uL{\mathrm{L}}

\def \uR{\mathrm{R}}
\def \uD{\mathrm{D}}
\def \uS{\mathrm{S}}
\def \uT{\mathrm{T}}
\def \uV{\mathrm{V}}
\def \rd{\partial}
\def \xv{\mbox{\boldmath$x$}}
\def \yv{\mbox{\boldmath$y$}}
\def \hH{\hat{H}}

\def \hA{\hat{A}}
\def \ha{\hat{a}}

\def \hV{\hat{V}}

\def \hU{\hat{U}}
\def \hO{\hat{O}}
\def \hT{\hat{T}}

\def \hS{\hat{S}}

\def \hW{\hat{W}}

\def \hOmega{\hat{\Omega}}

\DeclareFontFamily{OT1}{pzc}{}
\DeclareFontShape{OT1}{pzc}{m}{it}{<-> s * [1.10] pzcmi7t}{}
\DeclareMathAlphabet{\mathpzc}{OT1}{pzc}{m}{it}

\begin{document}
\title{Engineering Fast High-Fidelity Quantum Operations With Constrained Interactions}

\author{Thales Figueiredo Roque}
\affiliation{Max Planck Institute for the Science of Light, Staudtstraße 2, 91058 Erlangen, Germany}

\author{Aashish A. Clerk}
\affiliation{Pritzker School of Molecular  Engineering,  University  of  Chicago, 5640  South  Ellis  Avenue,  Chicago,  Illinois  60637,  U.S.A.}

\author{Hugo Ribeiro}
\affiliation{Max Planck Institute for the Science of Light, Staudtstraße 2, 91058 Erlangen, Germany}

\begin{abstract}
Understanding how to tailor quantum dynamics to achieve a desired evolution is a crucial problem in almost all quantum
technologies.  We present a very general method for designing high-efficiency control sequences that are always fully compatible
with experimental constraints on available interactions and their tunability.  Our approach reduces in the end to finding control
fields by solving a set of time-independent linear equations.  We illustrate our method by applying it to a number of
physically-relevant problems:~the strong-driving limit of a two-level system, fast squeezing in a parametrically driven
cavity, the leakage problem in transmon qubit gates, and the acceleration of SNAP gates in a qubit-cavity system.
\end{abstract}

\maketitle

\section{Introduction}
\label{sec:intro}

The success of any nascent quantum technology will ultimately be limited by our ability to manipulate relevant quantum states.
Finding the required time-dependent control fields that generate with high accuracy a desired unitary evolution is in general not
a trivial task: it is sufficient to consider a simple driven two-level system in the strong driving limit~\cite{fuchs2009} to find
an example of a complex control problem.  This generic problem becomes even more complicated when including realistic constraints:
~unavailable control fields, bandwidth and amplitude limitations, etc.  Finding new widely applicable methods to attack such
problems is thus highly desirable.   

There are of course many existing approaches to quantum control. Of these, the most ubiquitous is to exploit numerical algorithms
(see e.g.~\cite{khaneja2005,krotov1983,somloi1993,doria2011,glaser2015,machnes2018}) based on optimal quantum control theory
\cite{werschnik2007}.  The methods ultimately rely on the numerical optimization of an objective function, for example the
fidelity of a desired target state with the actual time-evolved state.  For many problems the effective landscape of the objective
function has many local minima, which can make it challenging to find the truly optimal protocol.  While methods to overcome these
limitations exist~\cite{kirkpatrick1983,swendsen1986,whitley1994,wenzel1999}, they become difficult to implement as the dimension
of the control space increases. An alternative approach is to use an analytical method to design effective protocols; control
pulses designed in this way could then be further improved by using them to seed a numerical optimal control algorithm.  Analytic
methods are however often system specific (see e.g.~\cite{motzoi2009,economou2015}),  or only work with a specific restricted
class of dynamics (for example methods based on shortcuts to adiabaticity, which are specific to protocols based on
adiabatic evolution~\cite{demirplak2003,demirplak2008,berry2009,ibanez2012,chen2012,baksic2016,ribeiro2019}). These approaches
are also generally impractical in systems with many degrees of freedom or sufficiently complex interactions.  

\begin{figure}[t!]
        \includegraphics[width=\columnwidth]{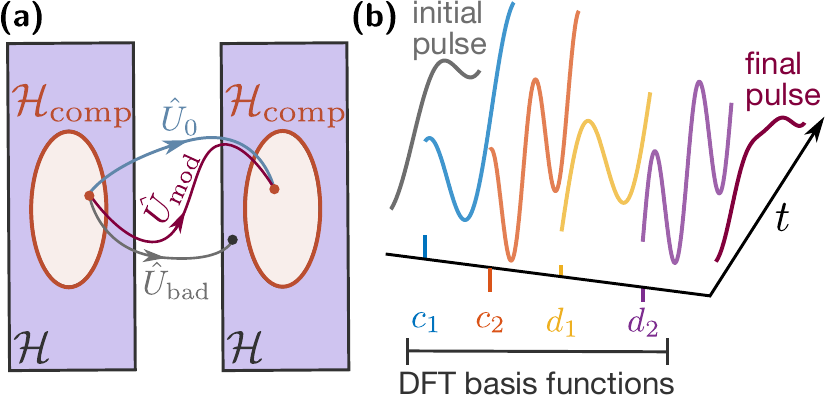}
	\caption{(Color Online) Generic quantum control problem. (a) An idealized unitary evolution
		($\hU_0$) maps an initial quantum state into a desired final state. The real evolution ($\hU$) does not
		allow one to reach the desired final state because it gets spoil by unwanted interactions neglected when deriving $\hU_0$.
		The effects of these interactions can be made arbitrary small by modifying the idealized control fields
		($\hU_\mm{mod}$). (b) The idealized control fields are modified by adding a finite-number of basis functions,
		e.g., discrete Fourier transform (DFT) basis functions, multiplied by free weights. The free weights are chosen
		such that the final pulse averages away the effects of the unwanted interactions.}
        \label{fig:fig01}
\end{figure}

In this work we present a general framework for constructing control fields that realize a desired evolution, in a manner that is
explicitly consistent with experimental constraints.  At its heart, it allows one to use the analytic solution of a simple control
problem to then find a high-fidelity pulse sequence for a more complex problem where a closed-form analytic solution is not
possible.  Our method has many potential virtues:  it is applicable to an extremely wide class of systems and protocols, produces
smooth control fields, and only requires one to numerically solve a finite set of linear equations.  It builds on the recently
proposed Magnus-based control method introduced in Ref.~\cite{ribeiro2017}, but greatly extends its power and applicability.     

Our generic goal is to use a specific time-dependent Hamiltonian $\hH (t)$ (whose form and tunability is constrained) to produce
(at time $t_\uf$) a desired unitary operation.  We start by splitting the Hamiltonian into two parts as $\hH (t) = \hH_0 (t) +\hV
(t)$, where $H_0(t)$ is simple enough to be analytically tractable, and $\hV (t)$ represents all the additional interactions that
make the problem unsolvable. The basic strategy then has two parts:
\begin{enumerate}[(1)]
    \item First, choose control fields in the ``simple" Hamiltonian $\hH_0(t)$ so that in the absence of $\hV(t)$, one realizes
	  the desired operation.  This can be done analytically.
    \item Adding back $\hV(t)$ will then destroy the ideal evolution.  We address this by modifying available control fields so as
	  to  average out the impact of $\hV(t)$.  This amounts to adding a control correction to the full Hamiltonian:  $\hH(t)
	  \rightarrow \hH(t) + \hW(t)$ [see Fig.~\ref{fig:fig01}~(a)]. 
\end{enumerate}

The question is of course how to find the desired control correction $\hW(t)$.  We address this using the strategy described
recently in Ref.~\cite{ribeiro2017}, where $\hW(t)$ is found perturbatively using a Magnus expansion \cite{magnus1954,blanes2009}.
A major limitation of this approach is that it often requires terms in $\hW(t)$ that are incompatible with the physical system at
hand (e.g.~interaction terms that do not exist, or that cannot be made time-dependent in the given experimental platform).  This
is where the present work makes a substantial contribution.  We introduce a novel way to find terms in the series expansion of
$\hW(t)$ that are \emph{always} compatible with all constraints.  We achieve this by expanding $\hW(t)$ at each order as a finite
sum of time-dependent basis functions multiplied by free weights.  Finding the required control corrections then amounts in most
cases to solving time-independent linear equations for these weights.

As we demonstrate through several examples, this methodology is both extremely flexible and effective; it can also work in systems
with many degrees of freedom.  The examples we consider include the strong non-RWA driving of a qubit (Sec.~\ref{sec:2LS}),
leakage errors in a superconducting qubit (Sec.~\ref{sec:transmon}), rapid squeezing generation in a parametrically driven bosonic
mode (Sec.~\ref{sec:DPA}), and accelerated SNAP gates~\cite{heeres2015,krastanov2015} in a coupled transmon-cavity system
(Sec.~\ref{sec:SNAP}).      

Note that the general idea of looking for control fields represented as a finite combination of basis functions was previously
used in Refs.~\cite{barends2014,martinis2014} to design two-qubit superconducting qubit gates that minimize leakage errors.  In
contrast to those works, our work is both more general and more systematic. Our approach is also complementary to a variational
approach for approximately finding STA protocols in complex systems that are compatible with experimental
constraints~\cite{sels2017,claeys2019}. 

\section{Theoretical Framework}
\label{sec:genth}

\subsection{Imperfect Unitary Evolution}

We consider the generic Hamiltonian:
\begin{equation}
	\hH (t) = \hH_0 (t) + \epsilon \hV (t).
	\label{eq:H}
\end{equation}
The Hamiltonian $\hH_0 (t)$ generates the desired time evolution, while $\hV (t)$ is the spurious ``error" Hamiltonian that disrupts the
ideal dynamics and which can be treated as a perturbation. The perturbative character of $\hV (t)$ can originate, e.g., from $\hV
(t)$ being proportional to a parameter $\epsilon \ll 1$, or because $\hV (t)$ is a fast oscillating function. In
Appendix~\ref{app:NonresonantPerturbations} we show why nonresonant error-Hamiltonians can also be corrected with the method presented
below. In this section, however, we consider the situation where $\hV (t)$ is proportional to a parameter $\epsilon \ll 1$ simply because
this allows one to count the orders of the perturbative series in a straightforward way. We stress, however, that one can apply the method
that we are about to introduce independently of the reason that makes $\hV (t)$ a perturbation. 

The time evolution operator generated by $\hH (t)$ is given by
\begin{equation}
	\hU (t) = \hU_0 (t) \hU_\uI (t).
	\label{eq:U}
\end{equation}
Here $\hU_0 (t)$ represents the ideal time evolution generated by $\hH_0 (t)$ ($\hbar=1$),
\begin{equation}
	\hU_0 (t)=\hT \exp \left[-i \int_0^t \di{t_1} \hH_0(t_1) \right],
	\label{eq:U0}
\end{equation}
where $\hT$ is the time ordering operator, and we assume that the time evolution starts at $t=0$. The effect of the error
Hamiltonian $\hV (t)$ on the dynamics is given by $\hU_\uI (t)$, which is defined as 
\begin{equation}
	\hU_\uI (t)=\hT \exp \left[-i \epsilon \int_0^t \di{t_1} \hV_\uI (t_1) \right].
	\label{eq:UI}
\end{equation}
Here, an operator $\hO (t)$ in the interaction picture is given by $\hO_\uI (t) = \hU_0^\dag (t) \hO (t) \hU_0 (t)$. 

Our goal is to have the time evolution operator at $t=t_\uf$  match a specific desired unitary operator $\hU_\uG$ ; the form of
the time evolution operator at earlier times is not relevant for us.  This is the case in many problems, the most prominent
example being the engineering of quantum gates. We also assume that $\hH_0 (t)$ provides us the desired time evolution at
$t=t_\uf$, i.e.  $\hU_0 (t_\uf) = \hU_\uG$. Consequently, the presence of a non-zero error Hamiltonian $\hV (t)$ disrupts the
evolution and prevents us to generate the desired evolution, since in general $\hU_\uI (t_\uf) \neq \mathbbm{1}$ [see
Eq.~\eqref{eq:U}]. 

\subsection{General Strategy to Correct Unitary Evolution}

To obtain the ideal unitary evolution at $t=t_\uf$, we wish to modify the time-dependence of $\hH (t)$ to cancel the deleterious
effects of $\hV (t)$. This is formally accomplished by introducing the modified Hamiltonian
\begin{equation}
	\hH_\mm{mod} (t) = \hH_0 (t) + \epsilon \hV (t) + \hW (t).
	\label{eq:Hmod}
\end{equation}
Here, $\hW (t)$ is an unknown control Hamiltonian that cancels, or at least mitigates, the effects of $\hV (t)$ on the
dynamics, bringing us closer to the desired time evolution [see Fig.~\ref{fig:fig01}~(a)]. The unitary evolution generated by
$\hH_\mm{mod} (t)$ is given by  $\hU_\mm{mod} (t) = \hU_0 (t) \hU_\mm{mod,\uI} (t)$, where 
\begin{equation}
	\hU_\mm{mod,\uI} (t) = \hT\exp \left[ -i \int_0^t \di{t_1} \hH_\mm{mod,\uI} (t_1) \right],
	\label{eq:Umod}	
\end{equation}
is the unitary evolution operator generated by the modified Hamiltonian in the interaction picture with respect to $\hH_0 (t)$. We
have 
\begin{equation}
	\hH_\mm{mod,\uI} (t) = \epsilon \hV_\uI (t) + \hW_\uI (t).
	\label{eq:HmodI}
\end{equation}
The desired unitary operator at $t=t_\uf$ is achieved if $\hU_\mm{mod,\uI} (t_\uf) = \mathbbm{1}$, i.e., $\hU_\mm{mod} (t_\uf) =
\hU_0 (t_\uf) = \hU_\uG$. 

A trivial solution to this problem is to take $\hW (t) = -\epsilon \hV(t)$.  This solution is almost always infeasible, as the
general form of $\hW(t)$ will be constrained by the kinds of interactions available in the system and their tunability.
Furthermore, we are only interested in  generating the correct unitary at $t=t_\uf$ and consequently cancelling the spurious
Hamiltonian at all times is in some sense demanding more than it is required. A better solution was found in
Ref.~\cite{ribeiro2017}, where one makes use of the fact that the time evolution at intermediate times is not important.  This
leads to relatively lax conditions that the control Hamiltonian $\hW (t)$ must satisfy. Nevertheless, finding an exact $\hW (t)$
is a complex task and generally one needs to resort to perturbation theory to find approximated solutions.

Let us start by writing $\hW (t)$ as a series in $\epsilon$,
\begin{equation}
	\hW (t) = \sum\limits_{n=1}^{\infty} \epsilon^n \hW^{(n)} (t).
	\label{eq:W}
\end{equation}
In order to find $\hW (t)$, one could work with the series expansion of the time-ordered exponential of Eq.~\eqref{eq:Umod}, but a
more convenient approach is to use the Magnus expansion~\cite{magnus1954,blanes2009}. With the Magnus expansion we can convert the
complicated time-ordered exponential to a simple exponential of an operator that can be expanded in a series:
\begin{equation}
	\hU_\mm{mod,\uI} (t) =  \exp\left[ \textstyle\sum\limits_{n=1}^{\infty} \hOmega_n(t) \right].
	\label{eq:Magnus}
\end{equation}
The terms of the Magnus expansion, $\hOmega_k(t)$, are recursively defined by differential equations~\cite{magnus1954,blanes2009},
with the first two terms being given by (see also Appendix~\ref{app:magnus})
\begin{align}
	\rd_t \hat{\Omega}_1(t) &= -i \hH_\mm{mod,\uI} (t),
	\label{eq:Magnus1}\\
	\rd_t \hat{\Omega}_2(t) &= \frac{1}{2} [\rd_t \hOmega_1(t), \hOmega_1(t)].
	\label{eq:Magnus2}
\end{align}
In order to correct the dynamics up to order $\mathcal{O}(\epsilon^m)$, one needs to find a control Hamiltonian $\hW (t)$ such
that $\hOmega_k (t_\uf) = \mathbf{0}$ for $k = 1, \ldots, m$. As shown in Ref.~\cite{ribeiro2017} this is accomplished if one
firstly truncates the series representing $\hW (t)$ [see Eq.~\eqref{eq:W}] up to order $m$ and then requires the operators
$\hW^{(n)}_\uI(t)$ to satisfy the following equations:
\begin{equation} 
	\epsilon^n \int_0^{t_\uf} \di{t} \ \hW^{(n)}_\uI (t) = -i \sum _{k=1}^n \hOmega_k^{(n-1)}(t_\uf),
	\label{eq:Wn}
\end{equation}
where $\hOmega_k^{(n)}(t)$ is the kth term of the Magnus expansion associated to the partially-corrected Hamiltonian 
\begin{equation}
	\hH_\mm{mod,\uI}^{(n)} (t) = \epsilon \hV_\uI(t) + \sum_{k=1}^n \epsilon^k \hW^{(k)}_\uI(t).
	\label{eq:HmodIk}
\end{equation}
Here, the series representing the correction $\hW (t)$ has been truncated at order $n$. 

To first order ($k=1$), Eq.~\eqref{eq:Wn} reduces to 
\begin{equation} 
	\int_0^{t_\uf} \di{t} \ \hW^{(1)}_\uI (t) = - \int_0^{t_\uf} \di{t} \ \hV_\uI(t).
	\label{eq:W1}
\end{equation} 
Equation~\eqref{eq:Wn} is the only restriction on the terms of the control Hamiltonian $\hW (t)$. This implies we have
considerable latitude in how we make our specific choice of $\hW (t)$. In what follows we fully exploit this freedom to
systematically find control Hamiltonians that are \textit{completely} compatible with experimental constraints on kinds and
tunability of available interactions.   

\subsection{Constrained Control Hamiltonians}

To proceed, we introduce a set of $N_\mm{op}$ time-independent Hermitian operators $\{\hA_j\}$ that form a basis for $\hH_0 (t)$,
$\hV (t)$, and $\hW (t)$.  By this, we mean that these operators allow for a unique decomposition of the different Hamiltonian
operators at each instant of time:
\begin{align}
	\hH_0(t) &= \sum\limits_j h_j (t) \hA_j,
	\label{eq:H0series} \\
	\hV (t) &= \sum\limits_j v_j (t) \hA_j,
	\label{eq:Vseries} \\
	\hW(t) &=  \sum\limits_j w_j (t) \hA_j.
	\label{eq:Wseries}
\end{align}
Here $h_j (t)$, $v_j (t)$, and $w_j (t)$ are the real control fields (expansion coefficients) associated with the decomposition of
$\hH_0 (t)$, $\hV (t)$, and $\hW (t)$, respectively. For instance, the elements of the set $\{\hA_j\}$ for a two-level system are
the Pauli operators $\hat{\sigma}_j$ with $j \in \{1,3\}$. We also introduce the Lie algebra $\mathfrak{g}$ generated by the set
of operators $\{-i \hA_j\}$ with the Lie bracket given by the commutation operation. Having a Lie algebra ensures that one can use
the basis formed by the set $\{\hA_j\}$ to decompose the operators generated by the Magnus expansion. Finally, we stress that
$N_\mm{op}$ can be finite even if the dimension of the Hilbert space is infinite. This is the case for quadratic bosonic forms
that can be characterized by the special unitary groups SU$(2)$ or SU$(1,1)$, which are associated to the Lie algebras su$(2)$ or
su$(1,1)$~\cite{yurke1986}.

Transforming Eqs.~\eqref{eq:Vseries} and \eqref{eq:Wseries} to the interaction picture defined by $\hH_0(t)$, we have 
\begin{equation}
	\hV_\uI (t) = \sum\limits_j v_j(t) \hA_{j, \uI}.
	\label{eq:VIseries0}
\end{equation}
Using the fact that $\{\hA_j\}$ forms a basis, we can write 
\begin{equation}
	\hA_{j, \uI} = \sum\limits_l a_{j,l}(t) \hA_l.
	\label{eq:Ajdecomp}
\end{equation}
Here, the functions $a_{j,l}(t)$ fully encode the action of the interaction picture transformation on our basis operators.

Substituting Eq.~\eqref{eq:Ajdecomp} in Eq.~\eqref{eq:VIseries0}, we obtain
\begin{equation}
	\hV_\uI(t) = \sum\limits_j \tilde{v}_j(t) \hA_j,
	\label{eq:VIseries}
\end{equation}
where we use tildes to denote control fields in the interaction picture, and we have
\begin{equation}
	\tilde{v}_j(t) = \sum\limits_l a_{l,j}(t) v_l(t).
	\label{eq:vtilde}
\end{equation}
Proceeding analogously for $\hW (t)$ and using the series representation defined in Eq.~\eqref{eq:W}, we get
\begin{equation}
	\hW_\uI^{(n)}(t) = \sum\limits_j \tilde{w}_j^{(n)}(t) \hA_j, 
	\label{eq:WIseries}
\end{equation}
with
\begin{equation}
	\tilde{w}_j^{(n)}(t) = \sum\limits_l a_{l,j}(t) w_l^{(n)}(t).
	\label{eq:wtilde}
\end{equation}

We now return to the fundamental equations of our approach, Eqs.~\eqref{eq:Wn}, which need to be satisfied to cancel the effects
of $\hV(t)$ to the desired order.  As written, these equations do not reflect any information about relevant experimental
constraints.  Typical examples of constraints are the inability to control the fields that couple to certain $\hA_j$, i.e.~that
particular field has to obey $w_j^{(n)}(t) = 0$.  Note that in general it is possible to have $h_j (t) \neq 0$ while one must work
with the condition  $w_j^{(n)}(t) = 0$.  Moreover, even if $w_j^{(n)}(t)$ can be controlled, it might have restrictions, e.g.
$w_j^{(n)}$ must be time independent or has bandwidth limitations. In the following we show how to derive equations for $w_j^{(n)}
(t)$ that obey Eqs.~\eqref{eq:Wn} and simultaneously fulfill the previously mentioned constraints. This then enables the design of
high fidelity control pulses that are \textit{fully} compatible with experimental constraints. As we discuss below, it is enough
to show how one derives equations for the first-order control fields $w_j^{(1)} (t)$, which must obey Eq.~\eqref{eq:W1}, since the
procedure for $w_j^{(n)} (t)$ is similar.

We proceed by substituting Eqs.~\eqref{eq:VIseries} and \eqref{eq:WIseries} into Eq.~\eqref{eq:W1}, which determines the
first-order correction Hamiltonian. We obtain an operator equation which can be split into $N_\mm{op}$ equations, one for each
operator $\hA_j$:
\begin{equation}
	\int_0^{t_\uf} \di{t} \tilde{w}_j^{(1)}(t) = -\int_0^{t_\uf} \di{t} \tilde{v}_j(t).
	\label{eq:CoeffsCorrGen}
\end{equation}
We stress that Eq.~\eqref{eq:CoeffsCorrGen} may be ill-defined since it is possible to have $\tilde{w}_j^{(1)}(t) = 0$ while
$\tilde{v}_j(t) \neq 0$ for certain values of $j$. We show in Sec.~\ref{sec:SNAP} how to deal with such situations for a large
class of problems. For the remainder of this section, we focus on the simpler case where we have a well-defined system of
equations.

The problem still remains of how to solve for $w_j^{(1)}(t)$; this is still a complex task since one is dealing with a system of
$N_\mm{op}$ coupled integral equations. This problem can be overcome by choosing an appropriate parametrization for the functions
$w_j^{(1)}(t)$. Here, since $w_j^{(1)}(t)$ must only have support on the interval $[0, t_\uf]$, we use a finite Fourier series
decomposition, 
\begin{equation}
	w_j^{(1)}(t) = \sum\limits_{k=0}^{k_\mm{max}} c_{jk}^{(1)} \cos\left(\omega_k t\right) + d_{jk}^{(1)}
	\sin\left(\omega_k t\right),
	\label{eq:Fourierw}
\end{equation}
with $\omega_k = 2 \pi k/t_\uf$ and $d_{j0}^{(1)} = 0$. This parametrization allows us to carry out the time integration over the
duration of the protocol and use the Fourier coefficients as the free parameters to satisfy the system of equations given by
Eq.~\eqref{eq:CoeffsCorrGen}. We stress that at this stage finding the first order correction that fulfills Eq.~\eqref{eq:W1} has
been reduced to determining a set of $N_\mm{coeffs} = N_\mm{op} \times (2 k_\mm{max} + 1)$ coefficients.  Note that one could use
other basis functions for the decomposition, e.g., Slepian functions~\cite{slepian1961}. 

The sum in Eq.~\eqref{eq:Fourierw} runs from $0$ to $k_\mm{max}$ which allows us to limit the bandwidth of the field associated to
$\hA_j$. We also note that $k_\mm{max}$ can take different values for different values of $j$. For constrained systems where a
particular field $w_j(t)$ must be time independent, we set all the coefficients in Eq.~\eqref{eq:Fourierw} to zero with the
exception of $c_{j0}$. If one requires $w_j^{(1)}(0) = w_j^{(1)}(t_\uf) =0$, then one finds using Eq.~\eqref{eq:Fourierw} that the
coefficients $c_{jk}^{(1)}$ must obey $\sum_{k=0}^{k_\mm{max}} c_{jk}^{(1)} =0$. For simplicity the summation in
Eq.~\eqref{eq:Fourierw} runs from $0$ to $k_\mm{max}$, but the more general case where the summation runs from $k_\mm{min}$ to
$k_\mm{max}$ is also allowed. 

We now can formulate the final basic equations of our approach.  We substitute Eqs.~\eqref{eq:wtilde} and \eqref{eq:Fourierw} in
the system of equations defined by Eq.~\eqref{eq:CoeffsCorrGen}.  Since we know the explicit time dependence of
$\tilde{w}_j^{(1)}(t)$, we can perform the time integration.  This leads to a system of time-independent $N_\mm{op}$ linear
equations than can be written in matrix form:
\begin{equation}
	M \: \xv^{(1)} = \yv^{(1)}.
	\label{eq:LinSysW1}
\end{equation}
Here, $\xv^{(1)}$ is a vector of coefficients (length $N_\mm{coeffs}$) determining the first order control correction that we are
trying to find.  In contrast, the matrix $M$ and the vector $\yv^{(1)}$ are known quantities: $\yv^{(1)}$ parameterizes the error
Hamiltonian $\hV(t)$, whereas $M$ encodes the dynamics of the ideal evolution generated by $\hH_0(t)$.

To be more explicit, the $\yv^{(1)}$ is a vector of length $N_\mm{op}$ whose components are the spurious error-Hamiltonian
elements we wish to average out, 
\begin{equation}
	y^{(1)}_j = - \int_0^{t_\uf} \di{t} \tilde{v}_j(t).
	\label{eq:yvector}
\end{equation}
$\xv^{(1)}$ is the vector of the $N_\mm{coeffs}$ unknown Fourier coefficients $c_{lk}^{(1)}$ and $d_{lk}^{(1)}$ that determine our
control corrections, c.f.~Eq.~(\ref{eq:Fourierw}).  We order these as follows
\begin{equation}
	x^{(1)}_j = 
	\begin{cases} 
		c_{lk}^{(1)} & \text{if } j \le j_0, \\
		d_{lk}^{(1)} & \text{if } j > j_0,
	\end{cases}
	\label{eq:xvector}
\end{equation}
where $j_0 = N_\mm{op}(k_\mm{max} + 1)$, and the indices $l$ and $k$ in Eq.~\eqref{eq:xvector} are functions of $j$. We have 
\begin{equation}
	l = 
	\begin{dcases} 
		(j-1)//(k_\mm{max} + 1) + 1 & \text{if } j \le j_0, \\
		(j - j_0 - 1)//k_\mm{max} + 1 & \text{if } j > j_0,
	\end{dcases}
	\label{eq:subindicel}
\end{equation}
and
\begin{equation}
	k = 
	\begin{dcases} 
		(j - 1) \% (k_\mm{max} + 1) & \text{if } j \le j_0, \\
		(j - j_0 - 1) \% k_\mm{max}  + 1 & \text{if } j > j_0.
	\end{dcases}
	\label{eq:subindicek}
\end{equation}
Here, $a//b$ denotes the integer division of $a$ by $b$ and $a\%b$ denotes the remainder of the integer division of $a$ by $b$.

Finally, $M$ is a $(N_\mm{op} \times N_\mm{coeffs})$ matrix that characterizes the evolution under the ideal Hamiltonian
$\hH_0(t)$. Recall that the interaction picture transformation generated by this Hamiltonian is described by the functions
$a_{jl}(t)$. The matrix elements of $M$ involve the Fourier series of these functions [see Eq.~\eqref{eq:Ajdecomp}]: 
\begin{equation}
	M_{ij} = 
	\begin{dcases} 
		\displaystyle\int_0^{t_\uf} \di{t} \ a_{l,i}(t) \cos(\omega_k t) & \text{if } j \le j_0, \\
		\displaystyle\int_0^{t_\uf} \di{t} \ a_{l,i}(t) \sin(\omega_k t) & \text{if } j > j_0,
	\end{dcases}
	\label{eq:Mmatrix}
\end{equation}
where $l$ and $k$ are given by Eqs.~\eqref{eq:subindicel} and \eqref{eq:subindicek}, respectively.  We stress that
Eqs.~\eqref{eq:yvector} to \eqref{eq:Mmatrix} are valid when the summation in Eq.~\eqref{eq:Fourierw} runs from $0$ to
$k_\mm{max}$ for all values of $j$, but they can be modified to describe other cases.

Higher orders controls are found with an identical procedure. Ultimately, each order is found by solving a system of time-independent
$N_\mm{op}$ linear equations similar to Eq.~\eqref{eq:LinSysW1} [see Appendix \ref{app:high_order_corr}]. 

In principle, a set of constrained controls fields that allows one to correct the dynamics up to order $n$, does not necessarily allows one
to correct the dynamics up to order $n+1$. In such situations, namely when the obtained linear system does not have a solution
\footnote{There are also situations where the linear system has a solution, however the solution coefficients are large in comparison with
the errors that they should correct. In such cases, the perturbative series of the correction Hamiltonian typically does not converge. We
treat this problem in Sec.~\ref{sec:transmon}.}, one usually has to choose another correction Hamiltonian for the system. There are
cases, however, where an alternative solution can be found. We illustrate this situation when we discuss the SNAP problem in
Sec.~\ref{sec:SNAP}.

\begin{table}[h!]
	\centering
	\begin{tabular}{l l l l} 
		\hline
		Symbol & Meaning & & Equation \\ [0.5ex] 
		\hline
		\hline
		$\hH_0(t)$ & Ideal Hamiltonian & & Eq.~\eqref{eq:H} \\ 
		$\hV(t)$ & Spurious ``error'' Hamiltonian & & Eq.~\eqref{eq:H} \\
		$\hW^{(k)}(t)$ & kth order correction Hamiltonian & & Eq.~\eqref{eq:W} \\
		$\hO_\uI$ & \simpletable{Operator in the interaction}{picture with respect to $\hH_0(t)$}  & & - \\
		$\hOmega_j(t)$ & \simpletable{$j$-th Magnus operator associated}{with $\hV_{\uI}(t)$} & & Eq.~\eqref{eq:Magnus} \\
		$\hOmega_j^{(k)}(t)$ & \simpletable{jth Magnus operator associated}{with the modified Hamiltonian} & & Eq.~\eqref{eq:HmodIk} \\
		$\hA_j$ & Basis operator of the Hilbert space & & - \\
		$v_j(t)$ & Decomposition coefficients of $\hV(t)$  & & Eq.~\eqref{eq:Vseries} \\ 
		$w_j^{(n)}(t)$ & Decomposition coefficients of $\hW^{(n)}(t)$ & & Eq.~\eqref{eq:Wseries} \\ 
		$a_{l, j}(t)$ & Decomposition coefficients of $\hA_{l, \uI}(t)$ & & Eq.~\eqref{eq:Ajdecomp} \\
		$\tilde{v}_j(t)$ & Decomposition coefficients of $\hV_{\uI}(t)$ & & Eq.~\eqref{eq:VIseries} \\  
		$\tilde{w}_j(t)$ & Decomposition coefficients of $\hW_{\uI}(t)$ & & Eq.~\eqref{eq:WIseries} \\ 
		$c_{jk}, d_{jk}$ & Fourier coefficients of $w_j(t)$ & & Eq.~\eqref{eq:Fourierw} \\ [1ex]
		\hline
	\end{tabular}
	\caption{Definition of the most important symbols.}
	\label{table:symbols}
\end{table}

\section{Applications}
\label{sec:Applications}

In this section, we apply our general strategy to several experimentally relevant problems. These examples highlight the fact that
our method is broadly applicable (without modification) to a wide range of very diverse problems.

\subsection{Strong Driving of a Two-Level System} 
\label{sec:2LS}

As a first example we consider the problem of a two-level system (qubit) in the strong driving limit. As we discuss below, this
regime generates a complex dynamics that renders precise control of the qubit hard to achieve. Several techniques were used to
predict control schemes which generate high-fidelity gates. Optimal control methods have been used, but the resulting control
fields are not bandwidth limited and cannot be accurately reproduced by an arbitrary wave form generator~\cite{scheuer2014}. An ad
hoc method based on time optimal control of a two-level system~\cite{boscain2006,garon2013} was also proposed:~it consists in
realizing Bang-Bang control with imperfect square control fields~\cite{hirose2018}. However, to achieve a gate with a reasonably
low error the imperfect square pulse must still have a relatively large bandwidth. A method based on analyzing the dynamics of the
system using Floquet theory has also been put forward~\cite{deng2015,deng2016}, but this transforms a low dimensional control
problem into a high dimensional one.

\begin{figure*}[t]
	\includegraphics[width=1.9\columnwidth]{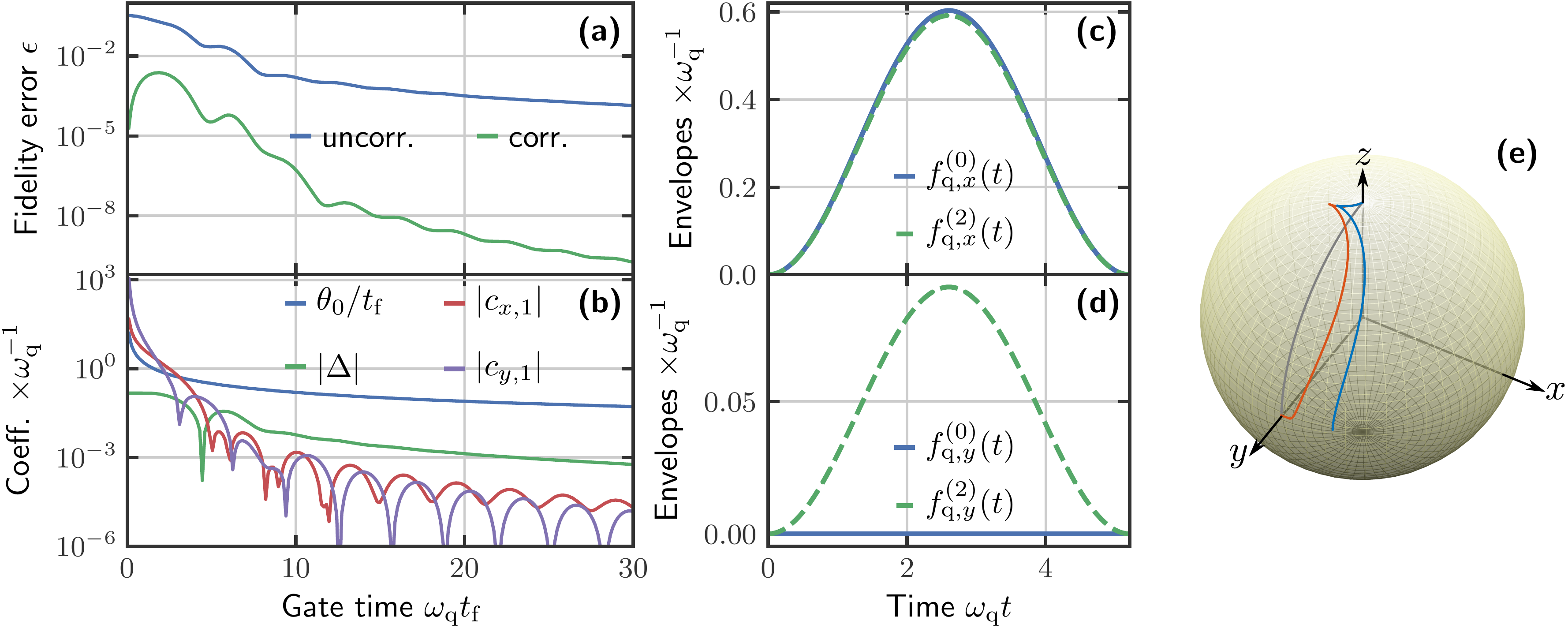}
	\caption{Modified dynamics of a qubit in the strong driving limit. (a) Average fidelity error for a Hadamard gate as a
	function of gate time. The blue trace is calculated using the uncorrected Hamiltonian [see Eq.~\eqref{eq:Hqubit}].
	The green trace is obtained for the modified Hamiltonian (up to second order). (b) Coefficient of the of original
	pulse $f_\uq(t)$ and the coefficients of the correction Hamiltonian as a function of the gate time. Here
	$c_{\alpha,1} = c_{\alpha,1}^{(1)} + c_{\alpha,2}^{(2)}$ and $\Delta = \Delta^{(1)} + \Delta^{(2)}$. (c-d)
	Original pulse and corrected pulse for $\omega_\ud t_f \approx 5$.  In (c) we plot $f_{\uq,x}^{(n)}(t)$ while in
	(d) we plot $f_{\uq,y}^{(n)}(t)$ [see Eq.~\eqref{eq:gen_pulse}]. (e) Trajectory on the Bloch sphere of the ideal
	dynamics (grey), of the uncorrected dynamics (blue), and of the corrected dynamics (orange).}
	\label{fig:fig02}
\end{figure*}

The Hamiltonian of a driven two-level system is given by
\begin{equation}
	\hH_\mm{qubit} (t) = \frac{\omega_\uq}{2} \s{z} + f_\uq(t) \cos(\omega_\ud t) \s{x},
	\label{eq:Hqubit}
\end{equation}
where $\omega_\uq$ is the qubit splitting frequency, $\omega_\ud$ is the driving frequency, $f_\uq(t)$ is the driving envelope, and
we introduce the Pauli operators:
\begin{equation}
	\begin{aligned}
		\s{x} &= \ketbra{0}{1}+\ketbra{1}{0},\\
		\s{y} &= i \ketbra{0}{1} - i \ketbra{1}{0},\\
		\s{z} &= \ketbra{1}{1}-\ketbra{0}{0}.
	\end{aligned}
	\label{eq:Paulimatrices}
\end{equation}
We label by $\ket{0}$ and $\ket{1}$ the ground and excited states of the system, respectively. We note that the Pauli
operators (multiplied by the imaginary number $-i$) define a Lie algebra with respect to the commutation operation.

In the weak driving limit, i.e. $f_\uq(t) \ll \omega_\ud\,\forall t$, Eq.~\eqref{eq:Hqubit} allows one to generate rotations
around the $x$-axis if one sets $\omega_\ud = \omega_\uq$. This is best understood in the frame rotating at the drive frequency
\footnote{We have used the unitary transformation $\hH_\mm{qubit} (t) \to \hH_\uR (t) = \hS_\ud^\dag (t) \hH_\mm{qubit} \hS_\ud
(t) - i \hS_\ud^\dag (t) \rd_t \hS_\ud (t)$ with $\hS_\ud (t) = \exp[-i \omega_\ud t \s{z}/2]$.}. In this frame the Hamiltonian is
given by $\hH_\uR (t) = \hH_{\uq,0} (t) + \hV_\uq(t)$ with 
\begin{equation}		
	\hH_{\uq,0} (t) = \frac{f_\uq(t)}{2} \s{x},
 	\label{eq:H0Qubit}
\end{equation}
and
\begin{equation}
	\hV_\uq (t) = v_{\uq,x}(t) \s{x} + v_{\uq,y}(t) \s{y}.
 	\label{eq:VbadQubit}
\end{equation}
The coefficients $v_{\uq,j}(t)$ are given by
\begin{equation}
	 \begin{aligned}		
		 v_{\uq,x} (t) &= \frac{f_\uq(t)}{2} \cos( 2 \omega_\ud t ), \\
		 v_{\uq,y} (t) &= -\frac{f_\uq(t)}{2} \sin( 2 \omega_\ud t ).
	 \end{aligned}
	\label{eq:vq}			
\end{equation}
Here, the driving is set on resonance with the qubit frequency, i.e., $\omega_\uq = \omega_\ud$. If the system is in the weak
driving limit, the fast oscillating terms (also known as counter-rotating terms) in $\hV_\uq(t)$ can be neglected as they average
themselves out over the long evolution time set by the slow varying envelope function $f_\uq(t)$. As a consequence, one can
approximate $\hH_\uR (t)$ by $\hH_{\uq,0} (t)$. This is known as the rotating wave approximation (RWA). The resulting Hamiltonian
generates a rotation of angle $\theta(t_\uf)$ around the $x$-axis, where we have introduced
\begin{equation}
	\theta(t) = \int_0^{t} \di{t_1} \ f_\uq(t_1).
	\label{eq:theta}
\end{equation}
However, when one deviates from the weak driving limit, the counter-rotating terms cannot be neglected anymore since they do not
average themselves out on short evolution times. As a result, the dynamics generated by $\hH_\uR (t)$ describes a complex rotation
around a time-dependent axis evolving in the $xy$-plane of an angle which is no more accurately described by
Eq.~\eqref{eq:theta}~\cite{bloch1940} [see Fig.~\ref{fig:fig02}~(e)]. To this day there is no known exact solution to this
problem, which makes the strong-driving limit impractical to control a qubit with high-fidelity. However, using the general
framework laid out in Sec.~\ref{sec:genth}, we can mitigate the effects of $ \hV_\uq (t)$ in situations where the RWA breaks down.
This allows us to generate any high-fidelity single-qubit gate beyond the RWA regime.

Given the constraints of the original problem, i.e., we only have temporal control over a field coupling to $\s{x}$ [see
Eq.~\eqref{eq:Hqubit}], we look for a correction of the form 
\begin{equation}
	\begin{aligned}
		\hW_\mm{qubit} (t) =& \sum\limits_n \left[ g_x^{(n)}(t) \cos(\omega_\ud t) + g_y^{(n)}(t) \sin(\omega_\ud t)
		\right] \s{x}.
	\end{aligned}
	\label{eq:wqubit}
\end{equation}
Here, $g_x^{(n)}(t)$ and $g_y^{(n)}(t)$ are unknown envelope functions. In addition to the driving field, we also have the liberty
to choose the driving frequency; nothing tells us that having $\omega_\ud = \omega_\uq$ is the best thing to do in terms of
control beyond the RWA. In the rotating frame, this is equivalent to have a non-zero detuning $\Delta = \omega_\uq - \omega_\ud$.
Therefore, we consider the following modified Hamiltonian in the rotating frame 
\begin{equation}
	\hH_{\uR,\mm{mod}} (t) = \hH_{\uq,0} (t)+ \hV_{\uq} (t) + \sum\limits_n \hW_\uq^{(n)}(t).
	\label{eq:Hr}
\end{equation} 
In terms of the Pauli operators, $\hW_\uq^{(n)}(t)$ is given by
\begin{equation}
	\hW_\uq^{(n)}(t) = w_{\uq,x}^{(n)}(t) \hat{\sigma}_x + w_{\uq,y}^{(n)}(t) \hat{\sigma}_y + w_{\uq,z}^{(n)}(t)
	\hat{\sigma}_z,
	\label{eq:Wq}
\end{equation}
with
\begin{equation}
	\begin{aligned}
		w_{\uq,x}^{(n)}(t) &= g^{(n)}(t) \cos(\omega_\ud t),\\
		w_{\uq,y}^{(n)}(t) &= -g^{(n)}(t) \sin(\omega_\ud t),\\
		w_{\uq,z}^{(n)}(t) &= \Delta^{(n)}. 
		\end{aligned}
	\label{eq:wq}
\end{equation}
In practice, having a control field with two-quadratures driving [see Eq.~\eqref{eq:wqubit}] and introducing a detuning has
given us the ability to implement $3$-axes control. We stress that there are other possible choices for $\hW (t)$, but they all
require more resources to be implemented experimentally [see Appendix \ref{app:DerivativeCorrection}]. Note that the modified
detuning is given by $\Delta=\sum_n \Delta^{(n)}$ in complete analogy to having the control fields represented by a series [see
Eq.~\eqref{eq:W}].

Following the general strategy presented in Sec.~\ref{sec:genth}, we first move to the interaction picture with respect to
$\hH_{\uq,0}(t)$ [see Eq.~\eqref{eq:H0Qubit}]. In the interaction picture, $\hV_{\uq}(t)$ [see Eq.~\eqref{eq:VbadQubit}] and the
control Hamiltonian $\hW_\uq^{(n)}(t)$ [see Eq.~\eqref{eq:Wq}] are respectively given by
\begin{equation}
	\hV_{\uq,\uI} (t) = \tilde{v}_{\uq,x}(t) \s{x} + \tilde{v}_{\uq,y}(t) \s{y} + \tilde{v}_{\uq,z}(t) \s{z},
	\label{eq:VqI}
\end{equation}
with
\begin{equation}
	\begin{aligned}
		\tilde{v}_{\uq,x}(t) &= \frac{f_\uq(t)}{2} \cos( 2 \omega_\ud t ),\\ 
		\tilde{v}_{\uq,y}(t) &= -\frac{f_\uq(t)}{2} \sin( 2 \omega_\ud t ) \cos{\theta},\\
		\tilde{v}_{\uq,z}(t) &= \frac{f_\uq(t)}{2} \sin( 2 \omega_\ud t ) \sin{\theta},
	\end{aligned}
	\label{eq:vqtilde}
\end{equation}
and
\begin{equation}
	\hW_{\uq,\uI}^{(n)}(t) = \tilde{w}_{\uq,x}^{(n)}(t) \s{x} + \tilde{w}_{\uq,y}^{(n)}(t) \s{y} + \tilde{w}_{\uq,z}^{(n)}(t) \s{z}
	\label{eq:WqI}
\end{equation}
with 
\begin{equation}
	\begin{aligned}
		\tilde{w}_{\uq,x}^{(n)}(t) &= g^{(n)}(t) \cos(\omega_\ud t),\\
		\tilde{w}_{\uq,y}^{(n)}(t) &= -g^{(n)}(t) \sin(\omega_\ud t)\cos{\theta}  + \Delta^{(n)} \sin{\theta},\\
		\tilde{w}_{\uq,z}^{(n)}(t) &= g^{(n)}(t) \sin(\omega_\ud t) \sin{\theta}  + \Delta^{(n)} \cos{\theta}.
	\end{aligned}
	\label{eq:wqtilde} 
\end{equation}
In Eqs.~\eqref{eq:vqtilde} and \eqref{eq:wqtilde}, we have omitted the explicit time dependence of $\theta$ for simplicity, i.e.,
$\theta = \theta(t)$ [see Eq.~\eqref{eq:theta}]. The next step consists in expanding the control fields
$\tilde{w}_{\uq,j}^{(n)}(t)$ ($j\in\{x,y,z\}$) [see Eq.\eqref{eq:wqtilde}] into a Fourier series. However, before proceeding it is
useful to notice the special form of the functions $\tilde{w}_{\uq,j}^{(n)}(t)$:~an unknown function that multiplies a known fast
oscillating function. It is therefore more suitable to just expand the unknown functions $g_{x}^{(n)}(t)$, $g_{y}^{(n)}(t)$ [see
Eq.~\eqref{eq:wqubit}], and $\Delta^{(n)}$ in a Fourier series and use the corresponding Fourier coefficients as the free
parameters to satisfy the system of equations generated by the Magnus-based approach. We stress, however, that one obtains exactly
the same results using the general procedure of Sec.~\ref{sec:genth} and imposing the necessary constraints on the Fourier series.

If we constrain $g_{\alpha=x,y}^{(n)}(t)$ to be zero at $t=0$ and $t=t_\uf$, which is often the case experimentally, we obtain the
following Fourier expansions
\begin{equation}
 	g_{\alpha}^{(n)}(t) = \sum\limits_{k=1}^{\infty} c_{\alpha, k}^{(n)} \left[ 1 - \cos(\omega_k t) \right] + d_{\alpha, k}^{(n)} \sin(\omega_k t), 
 	\label{eq:gjn}
\end{equation}
and
\begin{equation}
	\Delta^{(n)} = c_{z, 0}^{(n)} + \textstyle \sum\limits_{k=1}^{\infty} c_{z, k}^{(n)} \cos(\omega_k t) + d_{z, k}^{(n)} \sin(\omega_k t), 
 	\label{eq:deltan}
\end{equation}
where $\omega_k = 2 \pi/t_\uf$. Since we have a total of three equations of the form of Eq.~\eqref{eq:CoeffsCorrGen} to solve (one
for each Pauli operator), we need at least three free parameters. Consequently, we can set all coefficients to zero in
Eqs.~\eqref{eq:gjn} and \eqref{eq:deltan} except $c_{x, 1}^{(n)}$, $c_{y, 1}^{(n)}$ and $c_{z, 0}^{(n)}$~\footnote{We have chosen
this set of coefficients for simplicity. In principle one could choose another set of three coefficients.}. With this choice,
Eqs.~\eqref{eq:gjn} and \eqref{eq:deltan} reduce to 
\begin{equation}
	g_{\alpha=x,y}^{(n)}(t) = c_{\alpha, 1}^{(n)} \left[ 1 - \cos(\omega_1 t) \right],
 	\label{eq:gjn2}
\end{equation}
and
\begin{equation}
	\Delta^{(n)} = c_{z, 0}^{(n)}.
 	\label{eq:deltan2}
\end{equation}

The final step is to find the value of the free parameters $c_{x, 1}^{(n)}$, $c_{y, 1}^{(n)}$ and $\Delta^{(n)}$. We start by
verifying that by substituting Eqs.~\eqref{eq:gjn2} and \eqref{eq:deltan2} in Eq.~\eqref{eq:WqI}, we obtain a correction
Hamiltonian $\hW_{\uq,\uI}^{(n)}(t)$ [see Eq.~\eqref{eq:WqI}] that depends, as desired, linearly on the free parameters $c_{x,
1}^{(n)}$, $c_{y, 1}^{(n)}$ and $\Delta^{(n)}$. The system of equations defining the first order coefficients [$n=1$, see
Eq.~\eqref{eq:LinSysW1}], is given by
\begin{equation}
	P_\uq \xv_\uq^{(1)} = \yv_\uq^{(1)},
	\label{eq:LinSysQubit}
\end{equation}
where $\xv_\uq^{(1)} = \{ c_{x, 1}^{(1)}, c_{y, 1}^{(1)}, \Delta^{(1)} \}^T$ is the vector of unknown coefficients [see
Eq.~\eqref{eq:xvector}], $\yv_\uq^{(1)}= - \int_0^{t_\uf} \di{t} \{\tilde{v}_{\uq,x} (t), \tilde{v}_{\uq,y} (t), \tilde{v}_{\uq,z}
(t)\}^T$ is the vector of the spurious error-Hamiltonian elements with $\tilde{v}_{\uq,j} (t)$ ($j\in \{x,y,z\}$) defined in
Eq.~\eqref{eq:vqtilde}, and $P_\uq$~\footnote{The matrices $P_\uq$ and $M$ [see Eq.~\eqref{eq:Mmatrix}], although they fulfill the same
purpose, have different matrix elements. The difference arises because we are expanding in a Fourier series the unknown envelope
functions $g_\alpha^{(n)} (t)$ ($\alpha=x,y$) and the detuning $\Delta^{(n)}$ [see Eqs.~\eqref{eq:gjn} and \eqref{eq:deltan}]
instead of the functions $\tilde{w}_{\uq,j} (t)$ ($j\in \{x,y,z\})$) [see Eq.~\eqref{eq:wqtilde}].} is the matrix that
characterizes the evolution under the ideal Hamiltonian $\hH_{\uq,0} (t)$ [see Eq.~\eqref{eq:H0Qubit}].  The explicit matrix
elements of $P_\uq$ can be found in Appendix~\ref{app:LinearSystem2ls}. Higher-order correction Hamiltonians can be found in a similar
way.

In Fig.~\ref{fig:fig02}~(a), we plot the average fidelity error $\varepsilon$~\cite{pedersen2007} for a Hadamard gate generated
with an initial envelope
\begin{equation}
	f_\uq(t) = \frac{\theta_0}{t_\uf} \left[1 - \cos \left(\frac{2 \pi t}{t_\uf}\right)\right],
	\label{eq:fq}
\end{equation}	
with $\theta_0 = \pi/2$. Other gates can be realized by choosing $\theta_0 \in [0,2\pi]$. The blue trace shows the error for the
uncorrected evolution while the green trace shows the error of the corrected evolution up to second order. The latter, as one can
observe in Fig.~\ref{fig:fig02}~(a), globally increases when $\omega_\uq t_\uf$ decreases, but around $\omega_\uq t_\uf \simeq 1$
the error of the corrected evolution starts decreasing again. This can be understood by considering the limit $t_\uf \to 0$
($\omega_\uq t_\uf \to 0$). In this limit, we have  $\tilde{v}_{\uq,x} (t) \to f_\uq (t) /2$ and $\tilde{v}_{\uq,y} (t) =
\tilde{v}_{\uq,z} (t) \to 0$ [See Eq.~\eqref{eq:vqtilde}] which implies that $\hV_\uI (t)$ commutes with itself at all times. As a
consequence, one can find exact modifications to the control fields since only the first order of the Magnus expansion is
non-zero. However, as one can see in Fig.~\ref{fig:fig02}~(b), where we plot the coefficients of the correction versus the gate
time $t_\uf$, the modified control sequences require control fields with diverging amplitudes. Restricting ourselves to gate times
close to unity ($\omega_\uq t_\uf \simeq 1$), where the modified control sequences can be experimentally realized, our strategy
improves the error $\varepsilon$ by more than two orders of magnitude. In Figs.~\ref{fig:fig02}~(c) and (d), we compare the
original and corrected pulses for $\omega_\uq t_\uf \approx 5$. One can observe that the changes to the original pulse are small.
For convenience we write the nth order modified pulse as
\begin{equation}
	f_\mm{\uq, mod} (t) = f_{\uq,x}^{(n)}(t) \cos(\omega_\uq t) + f_{\uq,y}^{(n)}(t) \sin(\omega_\uq t),
	\label{eq:gen_pulse}
\end{equation}
where $f_{\uq,x}^{(n)}(t) = f_{\uq}(t) + \sum_{k=1}^n g_x^{(n)}(t)$ and $f_{\uq,y}^{(n)}(t) = \sum_{k=1}^n g_y^{(n)}(t)$. When $n
= 0$ we have simply the original pulse, thus $f_{\uq,x}^{(0)}(t) = f_{\uq}(t)$ and $f_{\uq,y}^{(0)}(t) = 0$.

\begin{figure}[t!]
	\includegraphics[width=\columnwidth]{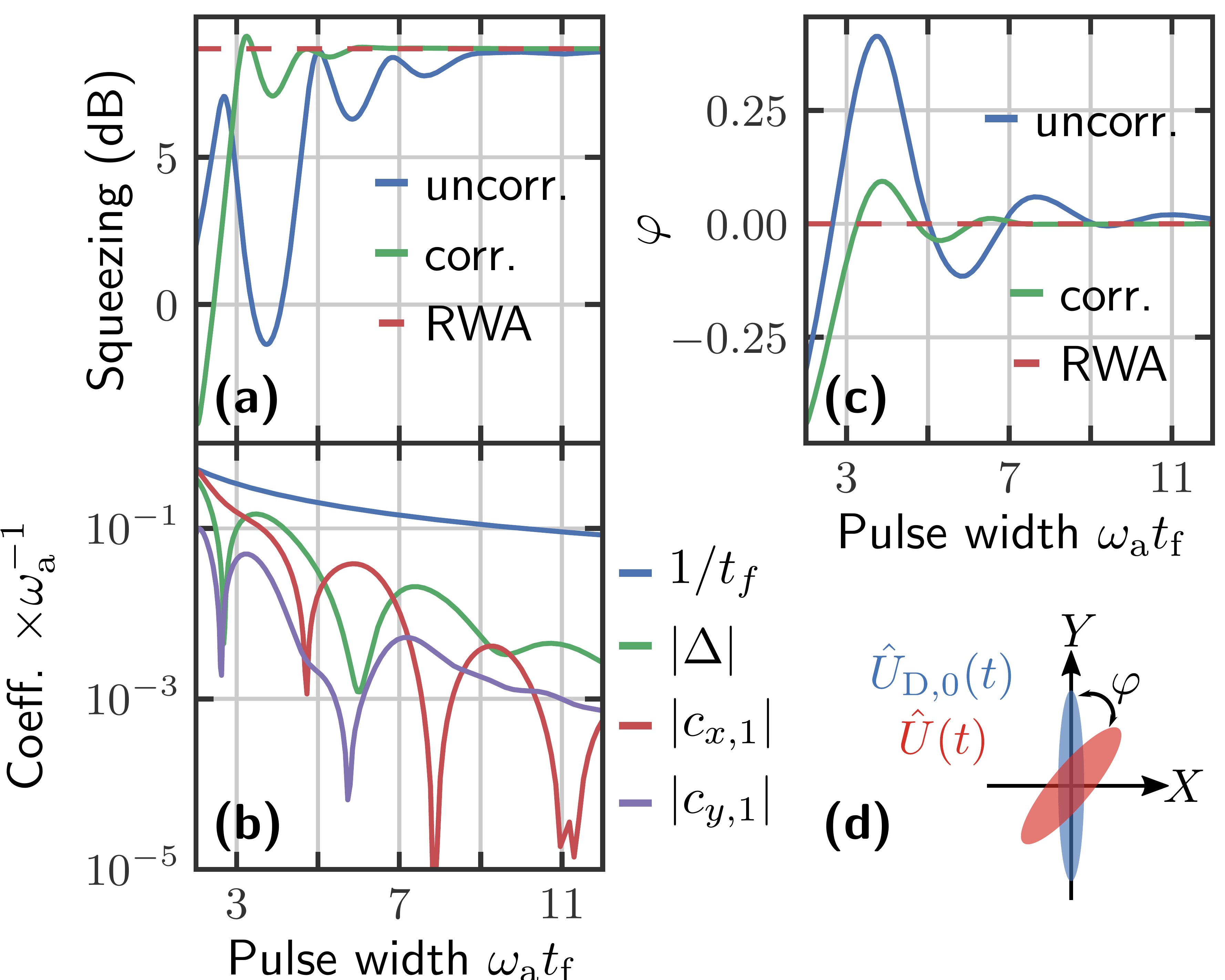}
	\caption{Fast generation of intra-cavity squeezing. (a) Intra-cavity squeezing as a function of the total evolution time.
	The red trace corresponds to the ideal case where the fast oscillating terms have been neglected. The blue trace
	shows the squeezing when the fast oscillating terms are present and no correction is used. The green trace shows
	the squeezing with the modified Hamiltonian (up to sixth order).  (Inset) Angle in the phase space where the
	squeezing is maximal as a function of gate time. The ideal case is $\Delta \varphi =0$. (b) Coefficients of the correction
	Hamiltonian as a function the total evolution time: $\{c_{x,1},\ c_{y,1},\ \Delta \} = \sum_{n=1}^3 \{c_{x,1}^{(n)},\
	c_{y,1}^{(n)},\ \Delta^{(n)} \}$.}
	\label{fig:fig03}
\end{figure}

\subsection{Strong driving of a Parametrically Driven Cavity} 
\label{sec:DPA}

As a second example, we consider the problem of fast generation of squeezed states using a parametrically driven cavity (PDC).
The ability to generate squeezed states with quantum oscillators is of particular interest since it allows one, among others,
to enhance sensing capabilities~\cite{burd2019} or to reach the single-photon strong coupling regime with optomechanical
systems using only linear resources~\cite{lemonde2016}. Recently, optimal control techniques have been used to achieve
squeezing of an optomechanical oscillator at finite temperature~\cite{basilewitsch2019}.

Here, we are interested in generating squeezing on a relatively short time scale by using a pulsed drive. As for the qubit problem
discussed in Sec.~\ref{sec:2LS}, this turns out to be a complex task due to fast counter-rotating terms that prevent the
preparation of the desired squeezed state.

The Hamiltonian of a PDC corresponds to having a harmonic oscillator with a modulated spring constant. This can be achieved, e.g.,
in the microwave regime by modulating the magnetic flux through a SQUID loop (flux-pumped Josephson parametric
amplifier)~\cite{ojanen2007,yamamoto2008}. We have
\begin{equation}
	\hH_\mm{PDC} (t) = \omega_\ua \ha^\dagger \ha + f_\uD(t) \sin(\omega_\ud t) \left(\ha + \ha^\dagger\right)^2,
	\label{eq:HDPA}
\end{equation} 
with $\ha$ ($\ha^\dagger$) the bosonic annihilation (creation) operator. The frequency of the mode $\ha$ is $\omega_\ua$ and the
drive has frequency $\omega_\ud$. 

It is convenient to introduce the operators~\cite{yurke1986} 
\begin{equation}
	\begin{aligned}
		\hat{\mu}_x &= \frac{1}{2} \left( \hat{a}^2 + \hat{a}^{\dagger 2}\right), \\
		\hat{\mu}_y &= -\frac{i}{2} \left( \hat{a}^2 - \hat{a}^{\dagger2} \right), \\
		\hat{\mu}_z &= \frac{1}{2} \left( \hat{a}^{\dagger} \hat{a} + \hat{a} \hat{a}^{\dagger} \right),
	\end{aligned}
	\label{eq:soperators}
\end{equation}
which define (multiplied by the imaginary number $-i$) a Lie algebra with respect to the commutation operation [see Appendix
\ref{app:DPA_operators}]. As mentioned earlier, since the Hamiltonian is quadratic, the three operators defined in
Eq.~\eqref{eq:soperators} are enough to completely describe the full dynamics in spite of having an infinite Hilbert space. The
action of these operators is best understood in the phase space defined by $\hat{x} = (\ha + \ha^{\dagger})/\sqrt{2}$ and $\hat{y}
= -i(\hat{a} - \hat{a}^{\dagger})/\sqrt{2}$:~$\hat{\mu}_x$ generates squeezing along the $x$-axis, $\hat{\mu}_y$ generates
squeezing along the $y$-axis, and $\hat{\mu}_z$ generates a rotation around the origin of the phase space.

In a frame rotating at a frequency $\omega_\ud/2 = \omega_\ua$, the Hamiltonian becomes $\hH_{\uD,R}(t) = \hH_{\uD,0}(t) +
\hV_{\uD}(t)$ with 
\begin{equation}
	\hH_{\uD,0}(t) =  f_\uD(t) \hat{\mu}_y,
	\label{eq:HD0}
\end{equation}
and
\begin{equation}
	\hV_{\uD}(t) = f_\uD(t) [ \sin(2 \omega_\ud t) \hat{\mu}_x - \cos(2 \omega_\ud t) \hat{\mu}_y + 2 \sin(\omega_\ud
	t) \hat{\mu}_z ].
	\label{eq:VD}
\end{equation}
In analogy with the qubit problem (see Sec.~\ref{sec:2LS}), one can neglect the fast oscillating Hamiltonian $\hV_{\uD}(t)$ [see
Eq.~\eqref{eq:VD}] in the weak driving limit (RWA), i.e., when $f_\uD(t) \ll \omega_\ud$ $\forall t$. This results in
$\hH_{\uD,R}(t) \approx \hH_{\uD,0}(t)$ and the generated dynamics corresponds to squeezing along the $y$-axis with a degree of
squeezing depending on $r(t_\uf)$, with 
\begin{equation}
	r(t) = \int_0^{t} \di{t_1} \ f_\uD(t_1).
 	\label{eq:F}
\end{equation}
As one deviates from the weak driving limit, $\hV_{\uD}(t)$ cannot be neglected anymore. The generated dynamics becomes then more
complex with the counter-rotating terms changing the direction along which the squeezing is generated as well as degrading the
final degree of squeezing [see Fig.~\ref{fig:fig03}~(d)].

To mitigate the effects of $\hV_{\uD}(t)$ [see Eq.~\eqref{eq:VD}], we consider a control Hamiltonian that corresponds to just
changing the initial form of the parametric modulation. This leads to the correction Hamiltonian 
\begin{equation}
	\begin{aligned}
	\hW_\mm{PDC} (t) = \sum\limits_n &\left[ g_x^{(n)}(t) \cos(\omega_\ud t) + g_y^{(n)}(t) \sin(\omega_\ud t) \right] \\
	&\times \left(\hat{a} + \hat{a}^{\dagger}\right)^2.
	\end{aligned}
	\label{eq:DPACorrDrive}
\end{equation}
Furthermore, we are at liberty to drive the PDC at a frequency that is detuned from that of mode $\ha$, 
\begin{equation}
	\frac{\omega_\ua}{2} - \omega_\ud = \Delta,
	\label{eq:DPACorrFreq}
\end{equation}
with $\Delta = \sum_n \Delta^{(n)}$ a static detuning.

Following the general procedure introduced in Sec.~\ref{sec:genth} (see Appendix \ref{app:DPA_Correction_Hamiltonian}), we can
easily determine $\Delta^{(n)}$, $g_x^{(n)}(t)$ and $g_y^{(n)}(t)$. We stress that in this example we correct the unitary
evolution generated by Eq.~\eqref{eq:HDPA}, which allows us to generate the ideal squeezing dynamics for any initial state. This
is in contrast to optimizing the dynamics to get optimal squeezing of the vacuum state only.  

In Fig.~\ref{fig:fig03}~(a), we plot the degree of squeezing $S$ as a function of the total evolution time $t_\uf$ for the RWA
(red trace), the uncorrected (blue trace), and the corrected (green trace) evolutions. The degree of squeezing is given by 
\begin{equation}
	S = -10 \log\left[\left( \langle \hat{y}^2 \rangle_f - \langle \hat{y} \rangle_f^2 \right)/\left(\langle \hat{y}^2 \rangle_i - \langle \hat{y} \rangle_i^2 \right)\right]
\end{equation}
where $\hat{y} = (\hat{a} - \hat{a}^{\dagger})/i \sqrt{2}$, and $\langle \hat{y} \rangle_{i, f} = \bra{\psi_{i, f}} \hat{y}
\ket{\psi_{i, f}}$ is the quantum average of the operator $\hat{y}$ with respect to the initial and final states, respectively.
Here, the initial state is the vacuum state $\ket{0}$. The initial pulse envelope is given by
\begin{equation}
	f_\uD(t) =\frac{1}{t_\uf}\left[1 - \cos\left(\frac{2 \pi t}{t_\uf}\right)\right].
	\label{eq:DPAInitEnv}
\end{equation}
Within the RWA the degree of squeezing is independent of the pulse width $t_\uf$, since the squeezing depends just on
$r(t_\uf)=c$. In the regime where the fast oscillating terms cannot be neglected, it is clear that the corrected evolution  gives
substantially better results (closer to the RWA evolution), specially for small values of $t_\uf$. In Fig.~\ref{fig:fig03}~(c), we
compute the deviation angle $\varphi$ in the phase space (with respect to the $y$-axis) where the maximum squeezing is obtained.
Ideally, the maximum squeezing should be in the direction of the $y$-axis and $\varphi$ should be zero.  With the correction
Hamiltonian $\varphi$ is much closer to the ideal value. In Fig.~\ref{fig:fig03}~(b), we plot the coefficients of the correction
Hamiltonian as a function of the total evolution time $t_\uf$. As for the qubit case, we observe that the modified control fields
can be seen as adding a small correction to the original control fields. 

\subsection{Transmon Qubit}
\label{sec:transmon}

\begin{figure*}[t]
	\includegraphics[width=2\columnwidth]{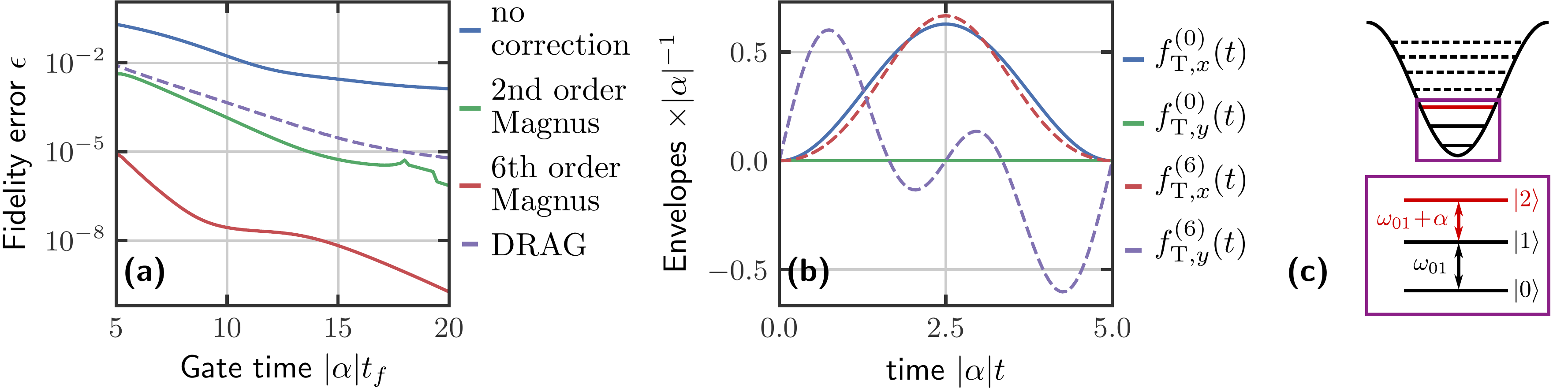}
	\caption{Fast and high-fidelity single qubit gates with a transmon. (a) Average fidelity error for a Hadamard gate as a
		function of the gate time. The blue trace is calculated using the uncorrected Hamiltonian [see
		Eq.~\eqref{eq:TLS}]. The green trace is obtained for the 2nd order corrected Hamiltonian. The red trace is
		obtained for the 6th order corrected Hamiltonian. The purple trace is obtained using the DRAG correction. (b)
		Initial envelope functions (solid lines) and 6th order corrected envelope functions (dashed lines) for
		$\abs{\alpha} t_\uf = 5$ [see Eq.~\eqref{eq:gen_pulse2}]. (c) Schematic energy level diagram of a transmon.}
	\label{fig:fig04}
\end{figure*}

As a next example we consider the problem of realizing single-qubit gates with a transmon qubit~\cite{koch2007}, where the logical
qubit states are encoded in the two lowest energy states of an anharmonic oscillator with eigenstates $\ket{n}$ [see
Fig.~\ref{fig:fig04}~(c)]. Since the oscillator is only weakly anharmonic, driving the $\ket{0} \leftrightarrow \ket{1}$
transition unavoidably leads to transitions to higher energy states outside of the computational subspace (leakage). Several
strategies have been put forward to suppress leakage while implementing a gate, with perhaps the most well-known approach being
DRAG (Derivative Removal by Adiabatic Gate)~\cite{motzoi2009,gambetta2011}. However, the correction predicted by DRAG cannot be
fully implemented experimentally as it also requires one to drive the $\ket{0} \leftrightarrow \ket{2}$ transition. There is no
charge matrix element connecting these states, hence it cannot be driven by an extra tone at the transition frequency. While
neglecting this unrealizable control field is the simplest thing to do, this is a somewhat uncontrolled approximation; further,
it has been demonstrated experimentally~\cite{chen2016} and theoretically~\cite{ribeiro2017} that this is indeed not the optimal
approach. In the rest of this section, we demonstrate how our general strategy allows one to systematically find control sequences
that are fully compatible with the constraints of the problem (i.e.~no direct $\ket{0} \leftrightarrow \ket{2}$ drive, no
time-dependent detuning), and also are highly efficient in suppressing both leakage and phase errors.  

As in the original DRAG paper, we consider the three-level Hamiltonian
\begin{equation}
    \begin{aligned}
	    \hat{H}_{\mm{TLS}}(t) =& \frac{\omega_\uT}{2} \s{z} + \left(\frac{3 \omega_\uT}{2}  + \alpha \right) \ketbra{2}{2} \\
	    & + f_\uT(t) \cos(\omega_\ud t) (\s{x} + \eta \n{x,1})
    \end{aligned}
    \label{eq:TLS}
\end{equation}
as an approximation of the weakly anharmonic oscillator. Here, $\omega_\uT$ is the frequency splitting between the energy levels
$\ket{0}$ and $\ket{1}$ while the frequency splitting between $\ket{1}$ and $\ket{2}$ is given by $\omega_\uT + \alpha$, where
$\alpha$ is the anharmonicity. We have also defined the operators
\begin{equation}
    \begin{aligned}
	\n{x,12} &= \ketbra{1}{2} + \ketbra{2}{1}, \ \n{y,12} = i\ketbra{2}{1} - i\ketbra{1}{2},  \\
	\n{x,02} &= \ketbra{0}{2} + \ketbra{2}{0}, \ \n{y,02} = i\ketbra{2}{0} - i\ketbra{0}{2},
\end{aligned}
    \label{eq:nuOperators}
\end{equation}
which describe transitions between the logical qubit states and the leakage state $\ket{2}$. These operators together with the
Pauli operators [see Eq.~\eqref{eq:Paulimatrices}] and the operator $\ketbra{2}{2}$ form the operator basis for this problem
[i.e.~the operators $\hA_j$ in Eq.~\eqref{eq:H0series}-\eqref{eq:Wseries}]. This set of eight operators (multiplied by the
imaginary number $-i$) also form a Lie algebra with respect to the commutation operation, thus this set of eight operators can
also be used to uniquely decompose the operators generated by the Magnus expansion. 

The control pulse consists of a drive at frequency $\omega_\ud$ and an envelope function $f_\uT(t)$. As one can see from
Eq.~\eqref{eq:TLS}, driving the $\ket{0} \leftrightarrow \ket{1}$ transition also results in the  $\ket{1} \leftrightarrow
\ket{2}$ being driven with a relative strength given by $\eta$, which unavoidably generates leakage out the qubit subspace. 

In a frame rotating with frequency $\omega_\ud$, the Hamiltonian is given by $\hH_\uT(t) = \hH_{\uT, 0}(t) + \hat{V}_{\uT}(t)$, where
\begin{equation}
	\hH_{\uT, 0}(t) = \alpha \ketbra{2}{2}  + \frac{f_\uT(t)}{2} \s{x},
	\label{eq:HT0}
\end{equation}
and
\begin{equation}
	\hat{V}_{\uT}(t) = \eta \frac{f_\uT(t)}{2} \n{x,1}.
	\label{eq:VT}
\end{equation}
Here, we assume that the driving is on resonance with the $\ket{0} \leftrightarrow \ket{1}$ transition, i.e., $\omega_\uT =
\omega_\ud$. The Hamiltonian $\hH_{\uT, 0}(t)$ gives us the desired interaction:~it couples the levels $\ket{0}$ and $\ket{1}$,
allowing one to perform unitary operations in the computational space, while leaving the level $\ket{2}$ isolated. The Hamiltonian
$\hat{V}_{\uT}(t)$ couples levels $\ket{1}$ and $\ket{2}$ leading to leakage out of the computational subspace. Note that we have
neglected the terms oscillating at frequencies close to $2 \omega_\ud$ in Eqs.~\eqref{eq:HT0} and \eqref{eq:VT} (RWA)\footnote{In contrast
to the examples treated in Secs.~\ref{sec:2LS} and \ref{sec:DPA}, counter-rotating terms are not a main source of error since
there is a relatively large separation between the driving frequency and the anharmonicity, i.e., $f_\uT(t)/\abs{\alpha} \gg f_\uT(t)/(2
\omega_\ud)$. As a result the error due to leakage out of the computational space is much larger than the error due to
counter-rotating terms. We stress that our framework would allow us to simultaneously deal with leakage and the counter-rotating
terms, but neglecting the latter  allows us to work with simpler expressions.}.

Given the constraints of the problem [see Eq.~\eqref{eq:HT0}], we want to find a correction that only involves modifying the drive
envelope we use, and possibly changing the detuning in a static manner. We thus write the control Hamiltonian (in the rotating
frame) as
\begin{equation}
	\begin{aligned}
		\hW_{\mm{TLS}}(t) =& \sum\limits_n \left[ g_x^{(n)}(t) \cos(\omega_\ud t) + g_y^{(n)}(t) \sin(\omega_\ud t)
		\right] \\
		&\times (\s{x}  + \eta \n{x,1}),
	\end{aligned}
	\label{eq:WTLS}
\end{equation}
with $g_x^{(n)}(t)$ and $g_y^{(n)}(t)$ the unknown envelope functions. Furthermore, we allow the drive frequency to be detuned
with respect to the base frequency of the transmon,
\begin{equation}
	\omega_\uT - \omega_\ud = \Delta.
	\label{eq:TransmonCorrFreq}
\end{equation}
As for the envelope functions, the detuning is parametrized as a series:~$\Delta = \sum_n \Delta^{(n)}$.

Within our framework, we would in principle need a total of eight free parameters to satisfy Eqs.~\eqref{eq:CoeffsCorrGen}, which
determine the first-order correction; this is because there are eight operators in the basis. Taking into account that $\ket{2}$,
which is outside the computational space, is of no interest to us, the equation associated to the operator $\ketbra{2}{2}$ can be
neglected. More generally, the equations originating from operators $\hA_j$ that act strictly outside of the computational space
do not need to be fulfilled, and one can simply neglect them to arrive at the relevant system of equations for the given order. 

We are therefore left with seven equations to fulfill, and we need at least seven coefficients. However, we cannot select seven
coefficients at random:~we must retain enough harmonics to ensure that $g_x^{(n)}(t)$ and $g_y^{(n)}(t)$ have a bandwidth
comparable to $\abs{\alpha}$; this is needed to easily correct leakage transitions. This was also identified in an
earlier work by Schutjens \textit{et al.}~\cite{schutjens2013}, which also aims at finding modified pulses to mitigate
leakage errors in a transmon. Their strategy consists in suppressing the spectral weight associated to leakage transitions
from the control fields. A systematic way of ensuring that the control fields have enough bandwidth is to keep more
non-zero coefficients than necessary in their Fourier expansion. This choice leads to an underdetermined linear system of
equations which can be solved using the Moore–Penrose pseudo-inverse~\cite{moore1920,bjerhammar1951,penrose1955} (see
Appendix~\ref{app:TransmonQubit}).

To show the performance of our strategy, we considered the situation where one wants to perform a Hadamard gate in the
computational subspace. In Fig.~\ref{fig:fig04}~(a), we plot the average fidelity error as a function of the gate time $t_\uf$. We
compare the results obtained in the absence of any correction (blue trace) with the results for a 2nd order Magnus-based
correction (green trace), a 6th order Magnus-based correction (red trace), and the DRAG correction (purple
trace)~\cite{motzoi2009}. The results show that the 6th order Magnus correction reduces the average fidelity error by more than
four orders of magnitude for small $\abs{\alpha t_\uf}$, greatly outperforming the DRAG correction. In Fig.~\ref{fig:fig04}~(b)
we compare the original and modified pulses for $|\alpha| t_\uf = 5$. 
For convenience we write the nth order modified pulse as
\begin{equation}
	f_\mm{\uT, mod} (t) = f_{\uT,x}^{(n)}(t) \cos(\omega_\uq t) + f_{\uT,y}^{(n)}(t) \sin(\omega_\uq t),
	\label{eq:gen_pulse2}
\end{equation}
where $f_{\uT,x}^{(n)}(t) = f_{\uT}(t) + \sum_{k=1}^n g_x^{(k)}(t)$ and $f_{\uT,y}^{(n)}(t) = \sum_{k=1}^n g_y^{(k)}(t)$. The case $n
= 0$ corresponds to the original pulse, i.e., $f_{\uT,x}^{(0)}(t) = f_{\uT}(t)$ and $f_{\uT,y}^{(0)}(t) = 0$.

A legitimate concern at this point is related to the possibility of realizing the pulses obtained with the Magnus formalism, since
arbitrary waveform generators (AWG) have bandwidth limitations. We remind the reader, however, that our method allows direct
control over the bandwidth of the pulse through truncation of the Fourier series. If a stricter limitation over the bandwidth of
the correction pulse is needed, one can make use of Lagrange multipliers to look for solutions of the linear system. As a rule of
thumb, the minimum requirement of our method is that the AWG bandwidth should approximately be comparable to or larger than the
anharmonicity $|\alpha|$.  

\subsection{SNAP Gates} 
\label{sec:SNAP}

We now turn to an example that combines both qubit and bosonic degrees of freedom. The general problem is to use a qubit coupled
dispersively to a cavity to achieve control over the bosonic cavity mode. A method for doing this was recently proposed and
implemented experimentally in a superconducting circuit QED architecture:~the so-called SNAP gates (selective number-dependent
arbitrary phase gates) combined with cavity displacements~\cite{heeres2015,krastanov2015}. Our goal will be to use
our general method to accelerate SNAP gates without degrading their overall fidelity.

An optimal control approach based on GRAPE has been used to accelerate the manipulation of the bosonic cavity
mode~\cite{heeres2017}. There is, however, a major advantage in using SNAP gates in combination with cavity
displacements:~the SNAP gate can be made robust against qubit errors~\cite{reinhold2019}, i.e., noise acting on the qubit
will not affect the quantum state of the cavity.

As we will see, this problem involves an interesting technical subtlety. When introducing our general method in
Sec.~\ref{sec:genth}, we stressed that it is crucial for the Hamiltonian $\hW_\uI (t)$ describing the modification of the control
fields to have terms involving \textit{all} of the basis operators $\hA_j$ appearing in the Magnus expansion of the unitary
evolution generated by the error-Hamiltonian $\hV_\uI (t)$. If this was not true, it would seemingly be impossible to correct
errors proportional to these basis operators.  Surprisingly, there are cases where this conclusion is overly pessimistic.  In
certain cases, one can still use a modified version of our Magnus-based strategy which uses an alternate method for finding an
appropriate $\hW(t)$.  As we show below, correcting SNAP gates is an example of this kind of situation.  The general price we pay
is that now, to find an appropriate set of control corrections, we need to solve a nonlinear set of equations (instead of the
linear equations in Eq.~(\ref{eq:LinSysW1}) that we used in all the previous examples).

The basic setup for SNAP gates involves a driven qubit that is dispersively coupled to a cavity mode.  The Hamiltonian is
$\hH_{\mathrm{SNAP}}(t) = \hH_{\uq\uc} + \hH_{\uD}(t)$, with
\begin{equation}
\hH_{\uq\uc} = \frac{1}{2}\left(\omega_\uq + \chi \ha^\dagger \ha \right) \s{z} + \omega_{\uc} \ha^{\dagger} \ha, 
	\label{eq:Hqc}
\end{equation}
and
\begin{equation}
	\hH_{\uD}(t) = [f_x(t) \cos(\omega_\ud t) + f_y(t) \sin(\omega_\ud t)] \s{x}.
	\label{eq:HSNAPd}
\end{equation}
The Pauli operators $\s{\alpha}$ act on the Hilbert space of the qubit and have been defined in Eq.~\eqref{eq:Paulimatrices}. We
also introduce the annihilation (creation) operator $\ha$ ($\ha^\dagger$) destroying (creating) an excitation of the oscillator.
The qubit is driven by two independent pulses, $f_x(t)$ and $f_y(t)$, which couple both to $\s{x}$ with the same frequency
$\omega_\ud$ but with different phases.

In the interaction picture with respect to $\hH_{\uq\uc}$, the Hamiltonian becomes
\begin{equation}
	\begin{aligned}
		\hH_\uS(t) &= \frac{1}{2} \sum_n  \Big( f_x(t) \left[ \cos(\delta\omega_n t) \s{x} - \sin(\delta\omega_n t) \s{y}
		\right]  \\
		&-f_y(t)  \left[ \sin(\delta\omega_n t) \s{x} + \cos(\delta\omega_n t) \s{y} \right] \Big) \ketbra{n}{n},
	\end{aligned}
	\label{eq:Hs}
\end{equation}
where $\delta\omega_n = \omega_{\uq} + \chi n - \omega_\ud$, $\ket{n}$ is a bosonic number state, and we have neglected fast
oscillating terms. If the drive is now chosen to fulfil $\omega_\ud = \omega_{\uq} + \chi n_0$, so that the drive is resonant for
a particular number-selected qubit transition, the Hamiltonian defined in Eq.~\eqref{eq:Hs} can be written as $\hH_\uS(t) =
\hH_{\uS,0}(t) + \hV_\uS(t)$.  Here
\begin{equation}
	\hH_{\uS,0}(t) = \frac{1}{2} \Big[ f_x(t) \s{x} - f_y(t) \s{y} 
	\Big] \ketbra{n_0}{n_0},
	\label{eq:H0snap}
\end{equation}
is the resonant part of the Hamiltonian defined in Eq.~\eqref{eq:Hs} and allows one to generate a unitary operation in the
subspace spanned by $\{ \ket{g, n_0}, \ket{e, n_0} \}$.  In contrast 
\begin{equation}
	\begin{aligned}
		\hV_\uS(t) =& \frac{1}{2} \sum_{n \neq n_0} \Big( f_x(t) \left[ \cos(\delta\omega_n t) \s{x} - \sin(\delta\omega_n t)
		\s{y} \right]  \\
		&  -f_y(t) \left[ \sin(\delta\omega_n t) \s{x} + \cos(\delta\omega_n t) \s{y} \right] \Big)
		\ketbra{n}{n}
	\end{aligned}
	\label{eq:Vsnap}
\end{equation}
is the non-resonant part of Eq.~\eqref{eq:Hs}. This error Hamiltonian is responsible for the unwanted dynamics in the subspace
spanned by $\{ \ket{g, n}, \ket{e, n} \}$, for $n \neq n_0$. While in principle the effects of $ \hV_\uS(t)$ on the dynamics
cannot be avoided, they are minimal in the weak-driving regime where $f_x(t), f_y(t) \ll \chi$. In this limit, we can use
$\hH_{\uS,0}(t)$ to generate a dynamics that imprints a phase on $\ket{n_0}$ while leaving all other states $\ket{n}$ ($n\neq
n_0$) unchanged. Our general goal will be to relax this weak-driving constraint, allowing for a faster overall gate.  

For concreteness, we assume that the qubit is initially in the state $\ket{g}$ and the driving pulses $f_x(t)$ and $f_y(t)$ are
chosen such that the qubit undergoes a cyclic evolution, i.e., the trajectory on the Bloch sphere encloses a finite solid angle
and at $t = t_\uf$ the state of the qubit is back to $\ket{g}$. This leads to the accumulation of a Berry phase $\gamma$ at
$t=t_\uf$ for the qubit which conditioned on the state of the cavity being $\ket{n_0}$. In other words, 
\begin{equation}
	\hU_{\uS, 0} (t_\uf) \ket{g, n} = 
	\begin{cases}
		e^{i \gamma} \ket{g, n}  & \text{if } n = n_0, \\
		\ket{g, n} & \text{if }  n \neq n_0,
	\end{cases}
	\label{eq:U0snap}
\end{equation}   
where $\hU_{\uS, 0} (t_\uf) = \hT \exp \left[ -i \int_0^{t_\uf} \di{t} \: \hH_{\uS,0} (t) \right]$ is the unitary evolution
generated by the ideal Hamiltonian in Eq.~\eqref{eq:H0snap}.  This approach can be generalized so that the ideal evolution yields
different qubit phase shifts for a set of different cavity photon numbers.  One simply replaces the driving Hamiltonian [see
Eq.~\eqref{eq:HSNAPd}] by
\begin{equation}
	\hH_{\ud}(t) = \sum_{n=0}^{N-1} [f_{x,n}(t) \cos(\omega_{\ud,n}  t) + f_{y,n}(t) \sin(\omega_{\ud,n} t)] \s{x},
	\label{eq:HSNAPd2}
\end{equation} 
where $\omega_{\ud,n} = \omega_{\uq} + \chi n$. The pulse envelopes $f_{x,n}(t)$ and $f_{y,n}(t)$ are chosen such that one gets
the desired phase in the $n$th energy level.

Of course, the above ideal evolution requires that $f_{x,n}(t), f_{y,n}(t) \ll \chi$, constraining the overall speed of the gate.
Without this assumption, the effects of the off-resonant error interaction given by the generalization of  $\hV_\uS(t)$
[c.f.~Eq.~\eqref{eq:Vsnap}] cannot be neglected, and will compromise the ideal SNAP gate evolution.  Again, our goal is to mitigate
these errors, allowing for faster gates.

In the following, we consider for simplicity the situation where one wants to imprint a phase on a single energy level of the
oscillator. The extension to the more general situation where one imprints arbitrary phases in different levels is
straightforward. We truncate the bosonic Hilbert space and work only within the subspace formed by the $N_\mm{trunc}$ first number
states. This procedure is justified by the fact that SNAP gates are typically used to manipulate ``kitten"
states~\cite{heeres2015,krastanov2015}, which are themselves restricted to a truncated subspace of the original bosonic Hilbert
space.

\begin{figure*}[t]
	\includegraphics[width=1.9\columnwidth]{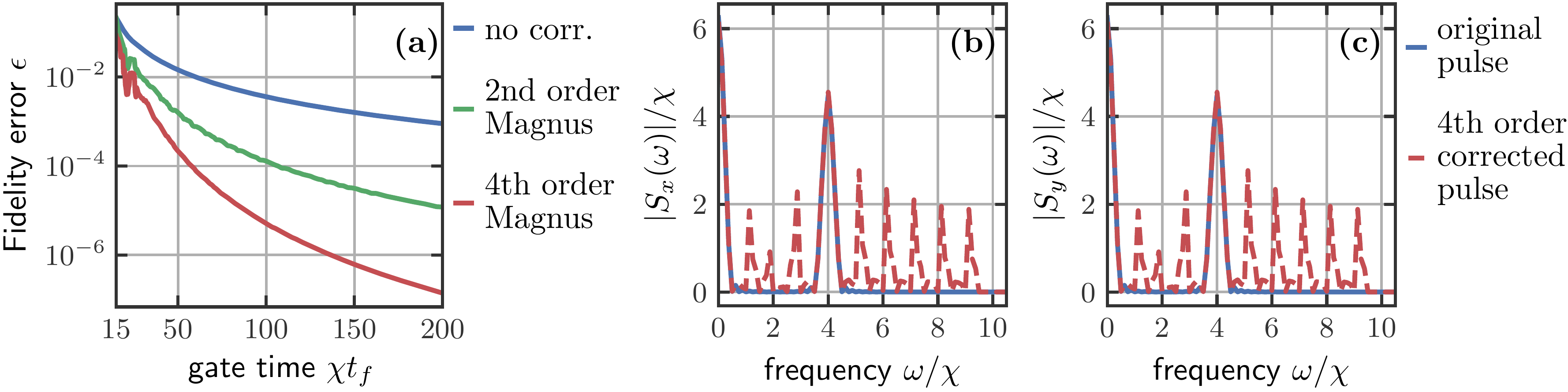}
	\caption{Accelerated SNAP gates. (a) Average fidelity error for a snap operation. A $\pi/2$ phase is imprinted in the
		cavity energy levels $\ket{0}$ and $\ket{4}$ simultaneously. The blue trace is calculated with the uncorrected
		Hamiltonian. The green trace is obtained with the 2nd order corrected Hamiltonian. The red trace is obtained with
		the 4th order corrected Hamiltonian.  (b-c) Spectrum of the $x$ and $y$ components of the original pulse envelope
		and of the 4th order corrected pulse envelope for $\chi t_\uf = 50$. The uncorrected envelope has peaks at $\omega
		= 0$ and $\omega = 4 \chi$. The corrected pulse has peaks close to $\omega = 0, \chi, \ldots, 9 \chi$. This means
		that the corrected pulse simply undoes residual rotations caused by the non-resonant interaction in the different
		bosonic-number-state subspaces in order to bring the final state close to the target state.
	}
	\label{fig:fig05}
\end{figure*}

As we did for the previous examples, we start by choosing a correction Hamiltonian $\hW_\mm{SNAP} (t)$ that one can realize
experimentally.  Here, this corresponds to a modification of the qubit drive amplitudes:
\begin{equation}
		\hW_{\mathrm{SNAP}}(t) \!=\! \sum\limits_{n=0}^{N-1} [g_{x,n}(t) \cos(\omega_{\ud,n} t) + g_{y,n}(t)
		\sin(\omega_{\ud,n} t)] \s{x}\\
	\label{eq:Wlabsnap}
\end{equation}
where $\omega_{\ud,n} = \omega_{\uq} + \chi n$. Moving to the interaction picture with respect to $\hH_{\uq\uc}$ [see
Eq.~\eqref{eq:Hqc}] and neglecting non-resonant terms, we obtain
\begin{equation}
	\hW_\uS(t) = \frac{1}{2} \sum\limits_{n=0}^{N-1} [g_{x,n}(t) \s{x} - g_{y,n}(t) \s{y}] \ketbra{n}{n}.
	\label{eq:WS}
\end{equation}

In the interaction picture defined by $\hH_{\uS,0}(t)$ [see Eq.~\eqref{eq:H0snap}], we find that the form of the
non-resonant error Hamiltonian is unchanged: 
\begin{equation}
	\hV_{\uS,\uI}(t) = \hV_\uS(t),
	\label{eq:VIS}
\end{equation}
since $\hH_{\uS,0}(t)$ commutes with $\hV_\uS(t)$; $\hH_{\uS,0}(t)$ and $\hV_\uS(t)$ act on orthogonal subspaces. On the other
hand, $\hW_\uS(t)$ acts on the whole Hilbert space, and is transformed when moving to the interaction picture.  We find:
\begin{equation}
	\begin{aligned}
		\hW_{\uS,\uI}&(t) = \hU_{\uS,0}^\dagger (t) \frac{1}{2} [g_{x,n_0}(t) \s{x} - g_{y,n_0}(t) \s{y}]
		\ketbra{n_0}{n_0} \hU_{\uS,0} (t) \\
		&+ \frac{1}{2} \sum\limits_{n=0}^{N-1} (1 - \delta_{n, n_0}) [g_{x,n}(t) \s{x} - g_{y,n}(t) \s{y}] \ketbra{n}{n}.
	\end{aligned}
	\label{eq:WIS}
\end{equation}
The first term of Eq.~\eqref{eq:WIS} acts on the $\{\ket{g,n_0}, \ket{e,n_0}\}$ subspace only and has terms proportional to all
three Pauli matrices. While the explicit expression is too lengthy to be displayed here, it can be readily found using the group
properties of the Pauli operators. The second term, which acts on the orthogonal subspace, has only terms proportional to $\s{x}
\ketbra{n}{n}$ and $\s{y} \ketbra{n}{n}$. This means that the correction Hamiltonian in Eq.~\eqref{eq:Wlabsnap} cannot correct
errors proportional to $\s{z} \ketbra{n}{n}$ (in the interaction picture) and which appear at 2nd order in the Magnus expansion of
$\hV_{\uS,\uI}(t)$ [see Eq.~\eqref{eq:VIS}]. Unfortunately, an analysis of the Magnus expansion generated by Eq.~\eqref{eq:VIS}
shows that these terms are by far the dominant source of errors which corrupt the ideal dynamics.  We are left with no choice but
to modify the general strategy of Sec.~\ref{sec:genth} that we have used successfully in all of the previous examples.  

The naive thing to do would be to find an alternative control Hamiltonian that directly provides terms proportional to $\s{z}
\ketbra{n}{n}$ in the interaction picture. However, in the lab frame this translates into a Hamiltonian with a dispersive coupling
constant dependent on photon number $n$, i.e., we would need a term $\sum_n \chi_n \ketbra{n}{n}$ in Eq.~\eqref{eq:Hqc}. This is
extremely difficult to achieve experimentally, hence we do not pursue this approach further.  

A more promising approach is to use the fact that even though our original (constrained) correction Hamiltonian $\hW_{\uS,\uI}(t)$
is missing important terms, these can nonetheless be dynamically generated. In the same way that $\hV_{\uS,\uI}(t)$ generates
problematic terms proportional to $\s{z} \ketbra{n}{n}$ at second order in the Magnus expansion, so can $\hW_{\uS,\uI}(t)$. Thus,
we look for a correction Hamiltonian $\hW^{(1)}(t)$ that cancels the \textit{sum} of the first two terms of the Magnus expansion:
\begin{equation}
	\hOmega_1^{(1)}(t_\uf) + \hOmega_2^{(1)}(t_\uf) = 0.
	\label{eq:MagnusSum0}
\end{equation}
This is in contrast of the general strategy in Sec.~\ref{sec:genth}, where we would just cancel the first term in the Magnus
expansion.  

We can use Eqs.~\eqref{eq:Magnus1} and \eqref{eq:Magnus2} to write Eq.~\eqref{eq:MagnusSum0} in terms of integrals involving
$\hV_{\uS,\uI}(t)$ [see Eq.~\eqref{eq:VIS}] and $\hW_{\uS,\uI}^{(1)}(t)$ [see Eq.~\eqref{eq:WIS} with $n=1$]. The explicit
equation can be found in the Appendix \ref{app:SNAPcorrection}. Here, to be able to correct for the term proportional to $\s{z}
\ketbra{n}{n}$, we cannot discard the higher order term generated by $\hW_{\uS,\uI}^{(1)}(t)$ in $\hOmega_2^{(1)}(t_\uf)$ (as the
standard procedure prescribes). Once this is done, we proceed as usual:~we expand the pulse envelopes $g_{x,n}(t)$ and
$g_{y,n}(t)$ in a Fourier series, and we truncate the series keeping a sufficiently large number of free parameters~\footnote{The
number of free parameters has a lower limit corresponding to the number of equations, but it is typically useful to have
more free parameters than equations. In such a case, one can use Lagrange multipliers to find solutions that minimize the sum of
the modulus squared of the free parameters.}. Following the strategy presented in Sec.~\ref{sec:genth}, we derive a system of
equations for the free parameters, but instead of obtaining a linear system of equations we get a system of quadratic equations
for the free parameters, since Eq.~\eqref{eq:MagnusSum0} is quadratic in $\hW_{\uS,\uI}^{(1)}(t)$. This system of equations can be
solved numerically; see Appendix \ref{app:SNAPcorrection}, where we also show how one can find higher order corrections.

In the situation where one wishes to imprint non-zero phases to all energy levels of the truncated Hilbert space, one can actually
solve the problem following the general (linear) strategy shown in Sec.~\ref{sec:genth}. In this case, since one is driving all
frequencies resonantly, the ideal unitary $\hU_{\uS,0}(t)$ acts on the whole truncated Hilbert space of the cavity. As a
consequence, transforming the correction Hamiltonian $\hW_\uS (t)$ [see Eq.~\eqref{eq:WS}] to the interaction picture will
generate terms proportional to $\s{z} \ketbra{n}{n}$ for all values of $n$ (see Appendix \ref{app:SNAPcorrection2}). We stress,
however, that SNAP gates are most often used to manipulate logical qubit states encoded in a finite superposition of same parity
bosonic number states~\cite{michael2016}, e.g., $\ket{0}_\uL = (\ket{0} + \ket{4})/\sqrt{2}$ and $\ket{1}_\uL = \ket{2}$.
Accelerating SNAP gates that act on such logical qubit states requires one to use the strategy that cancels the sum of the first
terms of the Magnus expansion [see Eq.~\eqref{eq:MagnusSum0}].

In Fig.~\ref{fig:fig05}~(a), we show the fidelity error when one tries to implement a fast SNAP gate that imprints a $\pi/2$ phase
in the cavity energy levels $\ket{0}$ and $\ket{4}$ simultaneously. This is similar to implement a $Z$-gate for a logical qubit
encoded in the states $\ket{0}_\uL = (\ket{0} + \ket{4})/\sqrt{2}$ and $\ket{1}_\uL = \ket{2}$. The envelope functions for $n = 0$ and $n=4$ are given by
\begin{equation}
	f_{x, n}(t) = 
	\begin{dcases}
	\frac{\pi}{2 t_\uf}\left[1 - \cos\left(4 \pi \frac{t}{t_\uf}\right)\right], \quad &t < \frac{t_\uf}{2}, \\
	0,  &t\geq \frac{t_\uf}{2},
	\end{dcases}
\end{equation}
and 
\begin{equation}
f_{y, n}(t) = 
	\begin{dcases}
	0, &t < \frac{t_\uf}{2}, \\
	\frac{\pi}{2 t_\uf} \left[1 - \cos\left(4 \pi \frac{t}{t_\uf}\right)\right], \quad &t\geq \frac{t_\uf}{2}.
	\end{dcases}
\end{equation}
For any other values of $n$, we have $f_{x, n}(t) = f_{y, n}(t) = 0$.  Here, we kept only ten energy levels for the cavity, i.e.,
the highest bosonic number state is $\ket{10}$, and the fidelity error was calculated using only the states within the truncated
Hilbert space.  We have plotted the fidelity as a function of gate time for the unmodified Hamiltonian (blue trace), for the
second-order (green trace), and for the fourth order (red trace) modified Hamiltonians. Since we are only manipulating the cavity
energy levels $\ket{0}$ and $\ket{4}$, we need to use the strategy that cancels the sum of the first terms of the Magnus expansion
[see Eq.~\eqref{eq:MagnusSum0}]. The fourth order modified Hamiltonian achieves fidelity errors that are at least one order of
magnitude smaller than the fidelity error of the original Hamiltonian. For larger values of $t_\uf$ the difference can reach
almost four orders of magnitude. 

In Fig.~\ref{fig:fig05}~(b) and (c) we show the spectrum of the original and modified pulses for a gate time of $\chi t_\uf = 50$.
The original pulse has only peaks located at $\omega = 0$ and $\omega = 4 \chi$, since these are the frequencies of the levels
being driven. The modified pulse, however, has peaks located at frequencies $\omega = 0, \chi, 2 \chi, \ldots, 9 \chi$. This shows
that the corrected pulse undoes residual rotations caused by the non-resonant interaction in the different bosonic number state
subspaces in order to bring the final state close to the target state. It is important to note that the modified pulse corrects
the dynamics only within the truncated Hilbert space. If the initial state of the cavity, i.e. the state before the SNAP operation
is performed, is not confined to the truncated Hilbert space, the corrected pulse will not bring any improvement in terms of
fidelity error, since the states lying outside the truncated Hilbert space will still be affected by the correction pulse.

\section{Conclusion} 

We have developed a method that allows one to design high-fidelity control protocols which are always fully compatible with
experimental constraints (available interactions and their tunability, bandwidth, etc.). At its core, our method uses the analytic
solution of a simple control problem as a starting point to solve perturbatively a more complex problem, for which it is
impossible to find closed-form analytic solutions. At the end of the day, the complex control problem is converted into solving a
simple linear system of equations. We have applied our method to a range of problems, including the leakage problem in a transmon
qubit and SNAP gates. We have shown how the control sequences predicted by our strategy allow one to substantially decrease the error
of unitary operations while simultaneously speeding up the time require to complete the protocols. Finally, we note that the
protocols generated by our method could be further improved by using them to seed a numerical optimal control algorithm. 

\section{Acknowledgements}

AC acknowledges partial support from the Center for Novel Pathways to Quantum Coherence in Materials, an Energy Frontier Research
Center funded by the Department of Energy, Office of Science, Basic Energy Sciences.

%



\begin{appendix}
\clearpage
\thispagestyle{empty}
\onecolumngrid
\begin{center}
\textbf{\large Appendices for: Engineering Fast High-Fidelity Quantum Operations With Constrained Interactions}
\end{center}

\section{Nonresonant Perturbations}
\label{app:NonresonantPerturbations}

In Sec.~\ref{sec:genth}, where we introduced the general strategy to correct unitary evolution, we have taken as a starting point
the Hamiltonian defined in Eq.~\eqref{eq:H}, where the spurious error-Hamiltonian $\hV(t)$ is multiplied by a small parameter
$\epsilon$, such that $\hV(t)$ can be considered as a perturbation. However, in all the examples discussed in the main text, the
spurious error-Hamiltonian is not multiplied by a small parameter $\epsilon$, but by an oscillating function of time. Since
integrating this oscillating function over a sufficiently long time interval leads to self-averaging, one can still view this
class of error-Hamiltonians as a perturbation. The Magnus expansion provides a good way to formally show that. 

Let us consider the generic oscillating Hamiltonian in the interaction picture,
\begin{equation}
	\hV_\uI(t) = \frac{1}{t_\uf}\left[ e^{-i \omega_\uV t} \hat{q}(t/t_\uf) + e^{i \omega_\uV t} \hat{q}^{\dagger}(t/t_\uf) \right],
	\label{eq:Vgen}
\end{equation}
where $\omega_{\uV}$ is the frequency at which the unwanted Hamiltonian $\hV_\uI(t)$ oscillates and
$\hat{q}(t/t_\uf)$ is a bounded operator. Considering the Magnus expansion of the evolution operator $\hU_\uI(t) = \hT
\exp[-i \int_0^{t}\di{t} \: \hV_\uI(t)]$, where $\hT$ is the time-ordering operator, we have to first order, at $t = t_\uf$,
\begin{align}
	\hOmega_1(t_\uf) &= \int_0^{t_\uf} \di{t} \: \hV_\uI(t)
	\label{eq:Omega1gen} \\
	&= \frac{1}{\omega_{\uV} t_{\uf}} \bigl[ i e^{-i \omega_\uV t} \hat{q}(t/t_\uf) \bigl|_0^{t_\uf} + H.c. \bigr] 
	+ \frac{1}{\omega_{\uV} t_{\uf}} \int_0^{t_\uf} \di{t} \: \bigl[ i e^{-i \omega_\uV t} \frac{ \mathrm{d}}{\di{t}} \hat{q}(t/t_\uf) 
	+ H.c. \bigr],
	\label{eq:Omega1gen2}
\end{align}
where Eq.~\eqref{eq:Omega1gen2} was obtained by integrating Eq.~\eqref{eq:Omega1gen} by parts. One can perform the integration by
parts repeatedly and obtain a series expansion for $\hOmega_1(t_\uf)$ in powers of $(\omega_{\uV} t_{\uf})^{-1}$. As one can see
from Eq.~\eqref{eq:Omega1gen2}, the leading order of $\hOmega_1(t_\uf)$ is $(\omega_{\uV} t_{\uf})^{-1}$. This suggests that
$(\omega_{\uV} t_{\uf})^{-1}$ plays the role of the small parameter $\epsilon$. Taking this analogy one step further, one would
naively assume that the leading order of $\hOmega_n(t_\uf)$ is $(\omega_{\uV} t_{\uf})^{-n}$, since this is the case when the
error-Hamiltonian is proportional to $\epsilon$. This is, however, not true for oscillating Hamiltonians [see
Eq.~\eqref{eq:Vgen}]. To show that, let us calculate the second order term of the Magnus expansion,
\begin{equation}
	\hOmega_2(t_\uf) = -\int_0^{t_\uf} \di{t_1} \int_0^{t_1} \di{t_2} \: [\hV_\uI(t_1), \hV_\uI(t_2)].
	\label{eq:Omega2gen}
\end{equation}
Substituting Eq.~\eqref{eq:Vgen} in Eq.~\eqref{eq:Omega2gen} and performing the integration in $t_2$ by parts, we obtain
\begin{equation}
	\hOmega_2(t_\uf) = -\frac{1}{t_\uf^2} \int_0^{t_\uf} \di{t_1} \: \bigl[ e^{-i \omega_\uV t_1} \hat{q}(t_1/t_\uf) + H.c.,
	-\frac{e^{-i \omega_\uV t_2}}{i \omega_\uV} \hat{q}(t_2/t_\uf)\bigr|_0^{t_1} + H.c. \bigr] + \mathcal{O}[(\omega_\uV
	t_\uf)^{-2}],
	\label{eq:Omega2gen2}
\end{equation}
where we have omitted terms that are $\mathcal{O}[(\omega_\uV t_\uf)^{-2}]$ and higher. Further simplifying
Eq.~\eqref{eq:Omega2gen2}, we find \footnote{Note that the terms calculated at $t_2 = 0$ give rise to terms proportional to
$(\omega_\uV t_\uf)^{-2}$. Therefore we omit these terms in Eq.~\eqref{eq:Omega2gen3}.}
\begin{align}
	\hOmega_2(t_\uf) &= -\frac{1}{t_\uf^2} \int_0^{t_\uf} \di{t_1} \: \Bigl\{ \frac{i}{\omega_\uV} \bigl[ \hat{q}(t_1/t_\uf),
	\hat{q}^{\dagger}(t_1/t_\uf) \bigr] + H.c. \Bigr\} + \mathcal{O}[(\omega_\uV t_\uf)^{-2}]
	\nonumber \\
	&= -\frac{1}{\omega_\uV t_\uf} \int_0^1 \di{x} \: \bigl\{ i \bigl[ \hat{q}(x),  \hat{q}^{\dagger}(x) \bigr] + H.c. \bigr\}
	+ \mathcal{O}[(\omega_\uV t_\uf)^{-2}],
	\label{eq:Omega2gen3}
\end{align}
which shows that $\hOmega_2(t_\uf)$ scales to leading order like $(\omega_\uV t_\uf)^{-1}$. 

Nevertheless, since we know the Magnus expansion converges, we must have that the leading order of $\hOmega_n(t)$ scales with
higher powers of $(\omega_\uV t_\uf)^{-1}$ for increasing $n$, but as we show above this dependence is not trivial.
Numerical tests suggest that for nonresonant perturbations the leading order of $\hOmega_j^{(0)}(t_\uf)$ is given by
\begin{align}
	\hOmega_j^{(0)}(t_\uf) =
	\begin{dcases}
		0 + \mathcal{O} \bigl[ (\omega_\uV t_\uf)^{-j/2} ] & \text{if } j \text{ is even}, \\
		0 + \mathcal{O} \bigl[ (\omega_\uV t_\uf)^{-(j+1)/2} ] & \text{if } j \text{ is odd}.
	\end{dcases}
	\label{eq:Omega_order}
\end{align}

Because of this property of nonresonant perturbations, one often needs to find the correction Hamiltonian $\hW (t)$ up to second
order to mitigate substantially the errors generated by $\hV (t)$. 

\section{The Magnus Expansion}
\label{app:magnus}

In the main text we showed the expression only for the first two terms of the Magnus expansion. For the bosonic system, however,
we have obtained a sixth order correction. The equations for the first four terms of the Magnus expansion are:
\begin{align}
	\partial_t \hat{\Omega}_1 =& -i \hH_\uI,
	\label{mes1}\\
	\partial_t \hat{\Omega}_2 =& -\frac{1}{2} [\hat{\Omega}_1, \partial_t \hat{\Omega}_1],
	\label{mes2}\\ 
	\partial_t \hat{\Omega}_3 =& -\frac{1}{2} [\hat{\Omega}_2, \partial_t \hat{\Omega}_1] 
	+ \frac{1}{12} [ \hat{\Omega}_1, [\hat{\Omega}_1, \partial_t \hat{\Omega}_1 ]],
	\label{mes3}\\	
	\partial_t \hat{\Omega}_4 =& -\frac{1}{2} [ \hat{\Omega}_3, \partial_t \hat{\Omega}_1] 
	+ \frac{1}{12} [ \hat{\Omega}_2, [\hat{\Omega}_1, \partial_t \hat{\Omega}_1]]
	+ \frac{1}{12} [ \hat{\Omega}_1, [\hat{\Omega}_2, \partial_t \hat{\Omega}_1]],
	\label{mes4}
\end{align}
A generator for arbitrary order terms, which is convenient to obtain high order terms, can be found in the subsection 2.3 of
Ref.~\cite{Sblanes2009}.

When trying to calculate the Magnus expansion terms, one might be tempted to calculate the terms iteratively, i.e. firstly
integrate eq.~(\ref{mes1}) to obtain $\hat{\Omega}_1$, then use this result to integrate Eq.~\eqref{mes2} and obtain
$\hat{\Omega}_2$, and so on. It is nonetheless much more efficient to treat all the terms that one intends to calculate as a
system of differential equations and solve them simultaneously. In this work we solved the differential equations using the
DifferentialEquations.jl package~\cite{Srackauckas2010} from the Julia programming language~\cite{Sbezanson2017}.	

\section{Arbitrary Order Corrections}
\label{app:high_order_corr}

A nth order correction, that generalizes Eq.~\eqref{eq:LinSysW1} of the main text, must satisfy the following
relation~\cite{Sribeiro2017}:
\begin{equation}
	\int_0^{t_\uf} \di{t} \ \hW_\uI^{(n)}(t) = -i \sum_{k=1}^{n} \hOmega_k^{(n-1)}(t_\uf).
	\label{eq:ArbMag}
\end{equation}
Since the set of operators $\{\hA_j\}$ forms a basis and $\{-i \hA_j\}$ generates a Lie algebra (see main text), we can write
$\hOmega_k^{(l)}$ as a linear combination of the operators $\{\hA_j\}$,
\begin{equation}
	\hat{\Omega}_k^{(l)}(t) = \textstyle \sum\limits_j \Omega_{k,j}^{(l)}(t) \hat{A}_j.
	\label{eq:OmegaExp}
\end{equation}
Substituting Eqs.~\eqref{eq:WIseries} of the main text and \eqref{eq:OmegaExp} in Eq.~\eqref{eq:ArbMag}, we obtain
\begin{equation}
	\int_0^{t_\uf} \di{t} \ \tilde{w}_j^{(n)}(t) = -i \sum_{k=1}^{n} \Omega_{k,j}^{(n-1)}(t_\uf) .
	\label{eq:CoeffsCorrGen2}
\end{equation}
The next steps are very similar to what was done for the first order correction. First, we expand $w_j^{(n)}(t)$ in a Fourier
series [see Eq.~\eqref{eq:Fourierw} of the main text]. Since $\tilde{w}_j^{(n)}(t) = \sum_l w_j^{(n)}(t) a_{l,j} (t)$, we can substitute
Eq.~\eqref{eq:Fourierw} of the main text in Eq.~\eqref{eq:CoeffsCorrGen2}, and we obtain
\begin{equation}
	M \: \xv^{(n)} = \yv^{(n)},
	\label{eq:LinSysWn}
\end{equation}
where $M$ is the same known $(N_\mm{op} \times N_\mm{coeffs})$ matrix obtained for $n=1$ [see Eq.~\eqref{eq:LinSysW1} of the main
text] and which encodes the dynamics of the ideal evolution generated $\hH_0 (t)$, $\xv^{(n)}$ is the vector of the
$N_\mm{coeffs}$ unknown Fourier coefficients $c_{lk}^{(n)}$ and $d_{lk}^{(n)}$ [see Eq.~\eqref{eq:Fourierw}], and $\yv^{(n)}$ is
the known vector of spurious elements we wish to average out. In the case where the summation in Eq.~\eqref{eq:Fourierw} runs from
$0$ to $k_\mm{max}$ for all values of $j$, the explicit expressions for the elements of the matrix $M$ are given by
Eq.~\eqref{eq:Mmatrix}. The elements of the vector $\xv^{(n)}$ are
\begin{equation}
	\xv^{(n)}_j = 
	\begin{dcases} 
		c_{l,k}^{(n)} & \text{if } j \le j_0, \\
		d_{l,k}^{(n)} & \text{if } j > j_0.
	\end{dcases}
\end{equation}
Here $j_0 = N_\mm{op}(k_\mm{max} + 1)$, and $l$ and $k$ are given by the Eqs.~\eqref{eq:subindicel} and \eqref{eq:subindicek} of
the main text. The elements of $\yv^{(n)}$ are given by
\begin{equation}
	\yv^{(n)}_j = -i  \sum_{k=1}^{n} \Omega_{k,j}^{(n-1)}(t_\uf).
\end{equation}	

\section{Strong Driving of a Two-Level System}

\subsection{Derivative-Based Correction}
\label{app:DerivativeCorrection}

While discussing the problem of strong driving of a two-level system, we mention that other choices for $\hW (t)$ were possible,
but that they all require more resources to be implemented experimentally. In this appendix, we illustrate this by considering the 
derivate-based control method introduced in Ref.~\cite{Sribeiro2017}.

Let us first summarize the principle on which the derivate-based control method is built on. According to Eq.~\eqref{eq:W1} of the
main text, the first order correction term must satisfy
\begin{equation}
	\int_0^{t_\uf} \di{t} \ \hat{V}_{\uq,\uI}(t) = - \int_0^{t_\uf} \di{t} \ \hat{W}_{I}^{(1)}(t) 
	\label{eq:tc2}
\end{equation}
where $\hat{V}_{\uq,\uI}(t)$ is given by Eq.~\eqref{eq:VqI} of the main text. Integrating the left hand side of Eq.~\eqref{eq:tc2}
by parts, we find
\begin{equation}
	\begin{aligned}
		\int_0^{t_\uf} \di{t} \ \hat{V}_{\uq,\uI}(t) =& \frac{f_\uq (t)}{4 \omega_\ud} \bigl[ \sin( 2 \omega_\ud t ) \s{x} + \cos( 2 \omega_\ud t ) 
		\cos{\theta} \s{y} - \cos( 2 \omega_\ud t ) \sin{\theta} \s{z} \bigr] \Bigl|_0^{t_\uf} \\
		&- \frac{1}{4 \omega_\ud} \int_0^{t_\uf} \di{t} \ \Bigl\{ \dot{f}_\uq (t) \sin( 2 \omega_\ud t ) \s{x} 
		+ \bigl[\dot{f}_\uq (t) \cos{\theta} - f_\uq^2(t)\sin{\theta} \bigr] \cos( 2 \omega_\ud t)  \s{y} \\
		&\qquad \qquad \qquad \quad + \bigl[\dot{f}_\uq (t) \sin{\theta} + f_\uq^2(t)\cos{\theta} \bigr] \cos( 2 \omega_\ud t)  
		\s{z} \Bigr\},
	\end{aligned}
	 \label{eq:tc3}
\end{equation}
where we have omitted the explicit time dependence of $\theta$ for simplicity, i.e., $\theta = \theta (t)$.
If the envelope function $f_\uq(t)$ vanishes at $t = 0$ and $t = t_\uf$, Eq.~\eqref{eq:tc3} becomes 
\begin{equation}
	\begin{aligned}
		\int_0^{t_\uf} \di{t} \ \hat{V}_{\uq,\uI}(t) =& 0
		- \frac{1}{4 \omega_\ud} \int_0^{t_\uf} \di{t} \ \Bigl\{ \dot{f}_\uq (t) \sin( 2 \omega_\ud t ) \s{x} 
		+ \bigl[\dot{f}_\uq (t) \cos{\theta} - f_\uq^2(t)\sin{\theta} \bigr] \cos( 2 \omega_\ud t)  \s{y} \\
		&\qquad \qquad \qquad \quad + \bigl[\dot{f}_\uq (t) \sin{\theta} + f_\uq^2(t)\cos{\theta} \bigr] \cos( 2 \omega_\ud t)  
		\s{z} \Bigr\}.
	\end{aligned}
	 \label{eq:tc4}
\end{equation}
Identifying Eq.~\eqref{eq:tc4} with Eq.~\eqref{eq:tc2}, we find that a possible choice for $\hW_{I}^{(1)}(t)$ is 
\begin{equation}
	\hW_\uI^{(1)}(t) = \tilde{w}_{1}^{(1)}(t) \s{x} + \tilde{w}_{2}^{(1)}(t) \s{y} + \tilde{w}_{3}^{(1)}(t) \s{z},
	\label{eq:tc5}
\end{equation}
where
\begin{align}
	\tilde{w}_{1}^{(1)}(t) &= \frac{1}{4 \omega_\ud} \dot{f}_\uq (t) \sin( 2 \omega_\ud t),
	\label{eq:tc6}\\
	\tilde{w}_{2}^{(1)}(t) &= \frac{1}{4 \omega_\ud} \bigl[\dot{f}(t) \cos{\theta} - f^2(t)\sin{\theta} \bigr] \cos( 2 \omega_\ud t),
	\label{eq:tc7}\\
	\tilde{w}_{3}^{(1)}(t) &= \frac{1}{4 \omega_\ud}\bigl[\dot{f}(t) \sin{\theta} + f^2(t)\cos{\theta} \bigr] \cos( 2 \omega_\ud t).
	\label{eq:tc8}
\end{align}

The second order control Hamiltonian $\hW_{I}^{(2)}(t)$ can be found by considering the Magnus expansion associated to the unitary
evolution generated by $\hH_\mm{mod,I}^{(1)} (t) = \hV_\uI (t) + \hW_\uI^{(1)} (t)$. By construction, the first term of the Magnus
expansion is given by 
\begin{equation}
	\hat{\Omega}_1^{(1)} (t) = -\frac{f_\uq (t)}{4 \omega_\ud} \bigl[ \sin( 2 \omega_\ud t ) \s{x} + \cos( 2 \omega_\ud t ) 
	\cos{\theta} \s{y} - \cos( 2 \omega_\ud t ) \sin{\theta} \s{z} \bigr],
	\label{eq:tc9}
\end{equation}
and vanishes at both $t=0$ and $t=t_\uf$ because $f_\uq (0) = f_\uq (t_\uf) = 0$.

According to Eq.~\eqref{eq:Magnus2} of the main text, the second term in the Magnus expansion is given by,
\begin{equation}
	\Omega_{2}^{(1)}(t) - \Omega_{2}^{(1)}(0) = -\frac{i}{2} \int_0^t \di{t}_1 \ \bigl[\hV_\uI (t_1) + \hW_\uI^{(1)} (t_1),
	\hat{\Omega}_1^{(1)} (t_1) \bigr].
	\label{eq:tc10}
\end{equation}
A natural choice for $\hW_{I}^{(2)}(t)$ is then (see Eq.~\eqref{eq:Wn} of the main text)
\begin{equation}
	\hat{W}_{I}^{(2)}(t) = -\frac{1}{2} \bigl[\hat{V}_\uI(t) + \hat{W}_{I}^{(1)}(t), \hat{\Omega}_1(t) \bigr],
\end{equation}

Transforming $\hW_{I}^{(1)}(t)$ and $\hW_{I}^{(2)}(t)$ back to the original frame, we obtain the correction Hamiltonian $\hW (t) =
\hW^{(1)}(t) + \hW^{(2)}(t)$. To implement this correction Hamiltonian, one would not only need to control in time the fields
coupling to $\s{x}$ and $\s{z}$, but one would also need an extra time-dependent control field that couples to $\s{y}$.

\subsection{Matrix Elements of $P_\uq$}
\label{app:LinearSystem2ls}

In this appendix, we give explicitly the matrix elements of the matrix $P_\uq$ introduced in Eq.~\eqref{eq:LinSysQubit} of the
main text when discussing the strong driving of a qubit. We have

\begin{align}
	P_{\uq,11} &= \int_0^{t_\uf} \di{t} \: [1 - \cos(\omega_1 t)] \cos^2(\omega_\ud t),
	\label{eq:ls2ls1}\\
	P_{\uq,12} &= \int_0^{t_\uf} \di{t} \: [1 - \cos(\omega_1 t)] \sin(\omega_\ud t) \cos(\omega_\ud t),
	\label{eq:ls2ls2}\\
	P_{\uq,13} &= 0,
	\label{eq:ls2ls3}\\
	P_{\uq,21} & = -\int_0^{t_\uf} \di{t} \: [1 - \cos(\omega_1 t)] \sin(\omega_\ud t) \cos(\omega_\ud t) \cos{\theta},
	\label{eq:ls2ls4}\\
	P_{\uq,22} & = -\int_0^{t_\uf} \di{t} \: [1 - \cos(\omega_1 t)] \sin^2(\omega_\ud t) \cos{\theta},
	\label{eq:ls2ls5}\\
	P_{\uq,23} & = \int_0^{t_\uf} \di{t} \: \sin{\theta},
	\label{eq:ls2ls6}\\
	P_{\uq,31} & = \int_0^{t_\uf} \di{t} \: [1 - \cos(\omega_1 t)] \sin(\omega_\ud t) \cos(\omega_\ud t) \sin{\theta},
	\label{eq:ls2ls7}\\
	P_{\uq,32} & = \int_0^{t_\uf} \di{t} \: [1 - \cos(\omega_1 t)] \sin^2(\omega_\ud t) \sin{\theta},
	\label{eq:ls2ls8}\\
	P_{\uq,33} & = \int_0^{t_\uf} \di{t} \: \cos{\theta},
	\label{eq:ls2ls9}
\end{align}
where once more we have omitted the explicit time dependence of $\theta$.

\section{Strong Driving of a Parametrically Driven Cavity}

\subsection{The Operators $\hat{\mu}_x$, $\hat{\mu}_y$, and $\hat{\mu}_z$}
\label{app:DPA_operators}

In this section, we give the commutation relations for the operators $\hat{\mu}_x$, $\hat{\mu}_y$, and $\hat{\mu}_z$ introduced in
Eq.~\eqref{eq:soperators} of the main text when discussing the strong driving of a parametrically driven cavity. These operators
behave as generators of the group SU$(1,1)$ and consequently generate the su$(1,1)$ Lie algebra, which one can readily verify by
computing the commutation relations. We have
\begin{equation}
	\begin{aligned}
	[\hat{\mu}_x, \hat{\mu}_y] &= 2 i \hat{\mu}_z, \\
	[\hat{\mu}_x, \hat{\mu}_z] &= 2 i \hat{\mu}_y, \\
	[\hat{\mu}_z, \hat{\mu}_y] &= 2 i \hat{\mu}_x. 
	\end{aligned}
\end{equation}
Therefore these three operators are enough to fully characterize the dynamics of the parametrically driven cavity in spite of
having an infinite Hilbert space. 

\subsection{Correction Hamiltonian}
\label{app:DPA_Correction_Hamiltonian}

In this appendix we give some more details about the steps of the general method applied to the problem of strong driving of a
parametrically driven cavity. 

Following the general procedure described in the general theory (see Sec.\ref{sec:genth} of the main text), we start by writing
the full modified Hamiltonian in the frame rotating at the drive frequency $\omega_\ud$:
\begin{equation}
	\hH_\mm{\uD, mod}(t) = \hH_{\uD,0}(t) + \hV_{\uD}(t) + \sum_n \hW_{\uD}^{(n)}(t),
\end{equation}
where $\hH_{\uD,0}(t)$ and $\hV_{\uD}(t)$ are, respectively, given by Eq.~\eqref{eq:HD0} and Eq.~\eqref{eq:VD} of the main text, and 
\begin{equation}
		\hW_{\uD}^{(n)}(t) = 2 g^{(n)}(t) \cos(\omega_\ud t)\hat{\mu}_x + 2 g^{(n)}(t) \sin(\omega_\ud t) \hat{\mu}_y + [\Delta^{(n)} + 2 g^{(n)}(t)] \hat{\mu}_z.  
	\label{eq:WDPA}
\end{equation}
Once more, we stress that the final detuning is given by $\Delta = \sum_n \Delta^{(n)}$. 

Following our recipe, we now move to the interaction picture with respect to $\hH_{\uD,0}(t)$. The Hamiltonian $\hV_{\uD}(t)$ is
then given by
\begin{equation}
	\hV_{\uD,\uI}(t) = \tilde{v}_{\uD,x}(t) \hat{\mu}_x + \tilde{v}_{\uD,y}(t) \hat{\mu}_y + \tilde{v}_{\uD,z}(t) \hat{\mu}_z,
	\label{eq:VDI}
\end{equation}
where
\begin{equation}
	\begin{aligned}
		\tilde{v}_{\uD,x}(t) &= f_\uD(t) [ \sin(2 \omega_\ud t) \cosh(2 r) + 2 \sin(\omega_\ud t) \sinh(2 r) ],\\
		\tilde{v}_{\uD,y}(t) &= - f_\uD(t) \cos(2 \omega_\ud t), \\
		\tilde{v}_{\uD,y}(t) &=  f_\uD(t) [ \sin(2 \omega_\ud t) \sinh(2 r) + 2 \sin(\omega_\ud t) \cosh(2 r) ].
	\end{aligned}
	\label{eq:vDtilde}
\end{equation}
Similarly, we find that the correction Hamiltonian in the interaction picture is given by
\begin{equation}
	\hW_{\uD,\uI}^{(n)}(t) = \tilde{w}_{\uD,x}^{(n)}(t) \hat{\mu}_x + \tilde{w}_{\uD,y}^{(n)}(t) \hat{\mu}_y + \tilde{w}_{\uD,z}^{(n)}(t) \hat{\mu}_z,	
	\label{eq:WDI}
\end{equation}
where
\begin{equation}
	\begin{aligned}
		\tilde{w}_{\uD,x}^{(n)}(t) &= 2 g^{(n)}(t) \cos(\omega_\ud t) \cosh(2 r) + [\Delta^{(n)} + 2 g^{(n)}(t)] \sinh(2 r), \\
		\tilde{w}_{\uD,y}^{(n)}(t) &= 2 g^{(n)}(t) \sin(\omega_\ud t), \\
		\tilde{w}_{\uD,z}^{(n)}(t) &= 2 g^{(n)}(t) \cos(\omega_\ud t) \sinh(2 r) + [\Delta^{(n)} + 2 g^{(n)}(t)] \cosh(2
		r).
	\end{aligned}
	\label{eq:wDtilde}
\end{equation}
For simplicity we have omitted the explicit time dependence of $r$, i.e.,  $r = r(t)$ [see Eq.~\eqref{eq:F} of the main text], in 
Eqs.~\eqref{eq:vDtilde} and \eqref{eq:wDtilde}.

\subsection{The Linear System of Equations}

In this section we give explicitly the matrix elements of $P_\uD$. As explained in Sec.~\ref{sec:2LS}, the matrix $P_\uD$ is
analog to the matrix $M$ introduced in Eq.~\eqref{eq:LinSysW1} of the main text and encodes the dynamics of the ideal evolution
generated by $\hH_{\uD,0} (t)$ (see Eq.~\eqref{eq:HD0} of the man text). The difference comes from our choice of expanding in a
Fourier series the unknown envelope functions $g_\alpha^{(n)} (t)$ ($\alpha =x,y$, see Eq.~\eqref{eq:DPACorrDrive} of the main
text) and the static detuning $\Delta^{(n)}$ instead of the functions $\tilde{w}_{\uD,\alpha} (t)$ [$\alpha =x,y,z$, see
Eq.~\eqref{eq:wDtilde}]. Within this framework, the system of linear equations that allows one to determine the coefficients
defining the nth order control correction is 
\begin{equation}
	P_\uD \xv_\uD^{(n)} = \yv_\uD^{(n)},
	\label{eq:LinSysPDC}
\end{equation}
where $\xv_\uD^{(n)} = \{ c_{x, n}^{(1)}, c_{y, n}^{(1)}, \Delta^{(n)} \}^T$ is the vector of unknown coefficients and $\yv_\uD^{(n)}$ is
the vector of spurious error-Hamiltonian elements. To first order we have $\yv_\uD^{(1)}= - \int_0^{t_\uf} \di{t} \{\tilde{v}_{\uD,x}
(t), \tilde{v}_{\uD,y} (t), \tilde{v}_{\uD,z}(t) \}^T$, with $\tilde{v}_{\uD,\alpha} (t)$ ($\alpha =x,y,z$) given in
Eq.~\eqref{eq:vDtilde}.

Finally, the matrix elements of $P_\uD$ are 
\begin{align}
	P_{\uD,11} &= \int_0^{t_\uf} \di{t} \: [1 - \cos(\omega_1 t)] \bigl\{[1 + \cos(4 t)] \cosh[2 r] + 2 \cos(2 t) \sinh[2 r] \bigr\},
	\label{lsDPA1}\\
	P_{\uD,12} &= \int_0^{t_\uf} \di{t} \: [1 - \cos(\omega_1 t)] \bigl\{ \sin(4 t) \cosh[2 r] + 2 \sin(2 t) \sinh[2 r] \bigr\},
	\label{lsDPA2}\\
	P_{\uD,13} &= \int_0^{t_\uf} \di{t} \sinh[2 r],
	\label{lsDPA3}\\
	P_{\uD,21} & = -\int_0^{t_\uf} \di{t} \: [1 - \cos(\omega_1 t)] \sin(4 t),
	\label{lsDPA4}\\
	P_{\uD,22} & = -\int_0^{t_\uf} \di{t} \: [1 - \cos(\omega_1 t)] [1 - \cos(4 t)],
	\label{lsDPA5}\\
	P_{\uD,23} & = 0,
	\label{lsDPA6}\\
	P_{\uD,31} & = \int_0^{t_\uf} \di{t} \: [1 - \cos(\omega_1 t)] \bigl \{ [1 + \cos(4 t)] \sinh[2 r] + 2 \cos(2 t) \cosh[2 r] \bigr\},
	\label{lsDPA7}\\
	P_{\uD,32} & = \int_0^{t_\uf} \di{t} \: [1 - \cos(\omega_1 t)] \bigl\{ \sin(4 t) \sinh[2 r] + 2 \sin(2 t) \cosh[2 r] \bigr\},
	\label{lsDPA8}\\
	P_{\uD,33} & = \int_0^{t_\uf} \di{t} \: \cosh[2 r],
	\label{lsDPA9}
\end{align}
where $\omega_1 = 2 \pi/t_\uf$ and once more we have omitted the explicit time dependence of $r$.

\section{Transmon Qubit}
\label{app:TransmonQubit}

\subsection{Interaction Picture}

In this appendix we show some steps of the general method applied to the transmon qubit that were omitted in the main text for
brevity. 

We first write the full modified Hamiltonian in a frame rotating with the drive frequency: 
\begin{equation}
	\hH_{\uT, \mm{mod}}(t) = \hH_{\uT, 0}(t) + \hat{V}_{\uT}(t) + \sum\limits_n \hat{W}_{\mm{T}}^{(n)}(t),
\end{equation}
where $\hH_{\uT, 0}(t)$ is given by Eq.~\eqref{eq:HT0}, $\hat{V}_{\uT}(t)$ is given by Eq.~\eqref{eq:VT}, and
\begin{equation}
        \hat{W}_{\uT}^{(n)} = \frac{1}{2} \Delta^{(n)} \bigl( \s{z} + 3 \ketbra{2}{2} \bigr) 
	+ \frac{1}{2} g_x^{(n)}(t) \left( \s{x} + \eta \n{x,12} \right) 
	+ \frac{1}{2} g_y^{(n)}(t) \left( -\s{y} + \eta \n{y,12} \right).	
	\label{eq:WT}
\end{equation}
As we previously did for the two-level system (see Sec.~\ref{sec:2LS}) and the DPA (see Sec.~\ref{sec:DPA}), we use the detuning
as yet another free parameter in the control Hamiltonian.

Before we move to the interaction picture with respect to $\hH_{\uT, 0}(t)$, let us adopt, for convenience, the following notation: 
\begin{equation}
	\begin{aligned}
		&\la{1}, \la{2}, \la{3}  = \s{x}, \s{y}, \s{z};  \\
		&\la{4}, \la{5}, \la{6}, \la{7} = \n{x,12}, \n{y,12}, \n{x,02}, \n{y,02}; \\
		&\la{8} = \ketbra{2}{2}.
	\end{aligned}
	\label{eq:lambdaops}
\end{equation}
Moving to the interaction picture with respect to $\hH_{\uT, 0}(t)$, the Hamiltonian $\hat{V}_{\uT}(t)$ is given by
\begin{equation}
	\hat{V}_{\uT, \uI}(t) = \tilde{v}_{\uT, 1}(t) \la{1} + \tilde{v}_{\uT, 2}(t) \la{2} + \tilde{v}_{\uT, 3}(t) \la{3} 
	+ \tilde{v}_{\uT, 4}(t) \la{4} + \tilde{v}_{\uT, 5}(t) \la{5} + \tilde{v}_{\uT, 6}(t) \la{6} 
	+ \tilde{v}_{\uT, 7}(t) \la{7} + \tilde{v}_{\uT, 8}(t) \la{8},
	\label{eq:VTI}
\end{equation}
where
\begin{equation}
	\begin{aligned}
		\tilde{v}_{\uT, 1}(t) &= \tilde{v}_{\uT, 2}(t) = \tilde{v}_{\uT, 3}(t) = 0, \\
	\tilde{v}_{\uT, 4}(t) &= \frac{\eta}{2} f_\uT(t) \cos(\theta/2) \cos(\alpha t), \\
	\tilde{v}_{\uT, 5}(t) &= \frac{\eta}{2} f_\uT(t) \cos(\theta/2) \sin(\alpha t), \\
	\tilde{v}_{\uT, 6}(t) &= \frac{\eta}{2} f_\uT(t) \sin(\theta/2) \sin(\alpha t), \\
	\tilde{v}_{\uT, 7}(t) &= -\frac{\eta}{2} f_\uT(t) \sin(\theta/2) \cos(\alpha t), \\
	\tilde{v}_{\uT, 8}(t) &= 0,
	\end{aligned}
\end{equation}
where for simplicity we have omitted the explicit time dependence of $\theta$, i.e.,
\begin{equation}
	\theta(t) = \int_0^t \di{t_1} f_\uT(t_1).
	\label{eq:theta2}
\end{equation}
Proceeding similarly we find 
\begin{equation}
	\hW_{\uT, \uI}^{(n)}(t) =  \tilde{w}_{\uT, 1}^{(n)}(t) \la{1} + \tilde{w}_{\uT, 2}^{(n)}(t) \la{2} 
	+ \tilde{w}_{\uT, 3}^{(n)}(t) \la{3} + \tilde{w}_{\uT, 4}^{(n)}(t) \la{4} + \tilde{w}_{\uT, 5}^{(n)}(t) \la{5} 
	+ \tilde{w}_{\uT, 6}^{(n)}(t) \la{6} + \tilde{w}_{\uT, 7}^{(n)}(t) \la{7}   + \tilde{w}_{\uT, 8}^{(n)}(t) \la{8}
	\label{eq:WTI}
\end{equation}
where 
\begin{equation}
	\begin{aligned}
		\tilde{w}_{\uT, 1}^{(n)}(t) &= \frac{1}{2} g_x^{(n)}(t), \\
		\tilde{w}_{\uT, 2}^{(n)}(t) &= \frac{1}{2} g_y^{(n)}(t) \cos{\theta} + \frac{1}{2} \Delta^{(n)} \sin{\theta}, \\
		\tilde{w}_{\uT, 3}^{(n)}(t) &= -\frac{1}{2} g_y^{(n)}(t) \sin{\theta} + \frac{1}{2} \Delta^{(n)} \cos{\theta}, \\
		\tilde{w}_{\uT, 4}^{(n)}(t) &= \frac{\eta}{2} \left[ g_x^{(n)}(t) \cos(\alpha t) \cos(\theta/2) - g_y^{(n)}(t)
		\sin(\alpha t) \cos(\theta/2) \right], \\
		\tilde{w}_{\uT, 5}^{(n)}(t) &= \frac{\eta}{2} \left[ g_x^{(n)}(t) \sin(\alpha t) \cos(\theta/2) + g_y^{(n)}(t)
		\cos(\alpha t) \cos(\theta/2) \right], \\
		\tilde{w}_{\uT, 6}^{(n)}(t) &= \frac{\eta}{2} \left[ g_x^{(n)}(t) \sin(\alpha t) \sin(\theta/2) + g_y^{(n)}(t)
		\cos(\alpha t) \sin(\theta/2) \right],\\
		\tilde{w}_{\uT, 7}^{(n)}(t) &= \frac{\eta}{2} \left[ - g_x^{(n)}(t) \cos(\alpha t) \sin(\theta/2) + g_y^{(n)}(t)
		\sin(\alpha t) \sin(\theta/2) \right],\\
		\tilde{w}_{\uT, 8}^{(n)}(t) &= \frac{3}{2} \Delta^{(n)}.
	\label{eq:wTtilde}
	\end{aligned}
\end{equation}

It is convenient to use the Gell-Mann $\hat{\lambda}$ operators to calculate commutators. The Gell-Mann $\hat{\lambda}$ operators
are given by Eq.~\eqref{eq:lambdaops}, except for $\la{8}$, which is given by
\begin{equation}
	\la{8} = (\ketbra{0}{0} + \ketbra{1}{1} - 2 \ketbra{2}{2})/\sqrt{3}
\end{equation} 
The Gell-Mann operators satisfy the following commutation relations:
\begin{equation}
	[\la{a}, \la{b}] = 2 i \textstyle \sum\limits_c f^{abc} \la{c},
\end{equation}
where the structure constants $f^{abc}$ are completely antisymmetric in the three indices, and are given by 
\begin{equation}
	f^{123} = 1, \ f^{147} = f^{165} = f^{246} = f^{257} = f^{345} = f^{376} = \frac{1}{2}, \ f^{458} = f^{678} = \frac{\sqrt{3}}{2}.
\end{equation}
The commutation relations of the Gell-Mann matrices are very convenient, specially when evaluating the Magnus expansion for this problem. 

\subsection{Choice of Free Parameters}

In the main text we showed that one has seven equations to fulfill for the transmon qubit problem, and this requires at least
seven free parameters. We also commented that it is important that the envelope functions $g_x^{(n)}(t)$ and $g_y^{(n)}(t)$ of the
correction Hamiltonian (see Eq.~\eqref{eq:WTLS} of the main text) have a bandwidth comparable to $\abs{\alpha}$, so that one can
access transitions between the levels $\ket{1}$ and $\ket{2}$. This becomes more clear if one considers the expressions of
$\tilde{w}_{\uT, j}^{(n)}(t)$ in Eqs.~\eqref{eq:wTtilde}. One can see that  $\tilde{w}_{\uT, j}^{(n)}(t)$ oscillates with
frequency $\abs{\alpha}$ for $j = 4, \ldots, 7$, while $\tilde{w}_{\uT, j}^{(n)}(t)$ is a slowly varying function for other values
of $j$. Since the effect of the correction Hamiltonian on the dynamics at $t = t_\uf$ is given by the integral of $\hat{W}_{\uT,
\uI}(t)$, the terms proportional to $\la{4}, \ldots, \la{7}$ average out unless $g^{(n)}(t)$ has a bandwidth comparable to
$\abs{\alpha}$. As a consequence $g_x^{(n)}(t)$ and $g_y^{(n)}(t)$ must have a bandwidth comparable to $\abs{\alpha}$.

Practically, this means that the envelope functions $g_x^{(n)}(t)$ and $g_y^{(n)}(t)$ associated to the correction Hamiltonian
(see Eq.~\eqref{eq:WTLS} of the main text) need to have a certain number of non-zero coefficients such that the condition on the
bandwidth can be satisfied. A systematic way of determining which coefficients are non-zero is to choose the coefficients of the
harmonics between $k=1$ and $k \simeq \abs{\alpha} t_\uf / 2 \pi$ in the Fourier expansion of the envelopes to be non-zero and set
all the other coefficients to zero. Furthermore, assuming that the detuning is time-independent, all coefficients of its Fourier
series except $c_{z,0}$ are zero. This typically gives us more than seven free coefficients in total, and we end up with an
underdetermined system of linear equations.

As mentioned in the main text, we can use the Moore–Penrose pseudo-inverse~\cite{Smoore1920,Sbjerhammar1951,Spenrose1955} to solve
this underdetermined system of linear equations. Importantly, the pseudo-inverse always exists, which guarantees that the linear
system always has a solution, and the pseudo-inverse also enforces that the solution has the smallest possible norm, which results
in having only a fraction of the free coefficients to actually be non-zero.

\section{SNAP Gates}

\subsection{Correction Hamiltonian}
\label{app:SNAPcorrection}

As discussed in the main text, the correction Hamiltonian for SNAP gates (see Eq.~\eqref{eq:Wlabsnap} of the main text) does not
allow one to correct terms proportional to $\s{z} \ketbra{n}{n}$ using the general strategy presented in Sec.~\ref{sec:genth}. As
we argue in the main text, the most important source of errors are precisely those originating from terms in the error-Hamiltonian
proportional to $\s{z} \ketbra{n}{n}$. This makes it necessary to find another strategy to correct those errors. 

A correction Hamiltonian with terms proportional to $\s{z} \ketbra{n}{n}$ in the rotating frame is out of question, since it would
require a dispersive coupling constant dependent on $n$. We must, therefore, abandon the general strategy used so far and look
for an alternative approach. 

Let us write explicitly the Magnus expansion, up to the second order, of the evolution operator associated to the modified
Hamiltonian $\hH_{mod,\uI}^{(1)}(t) = \hV_{\uS,\uI}(t) + \hW_{\uS,\uI}^{(1)}(t)$. We have 
\begin{equation}
	\begin{aligned}
		\hOmega_1^{(1)}(t_\uf) + \hOmega_2^{(1)}(t_\uf) =& \hOmega_1^{(0)}(t_\uf) + \hOmega_2^{(0)}(t_\uf) - i
		\int_0^{t_\uf} \di{t_1} \hW_{\uS,\uI}^{(1)}(t_1) \\
		& - \frac{1}{2} \int_0^{t_\uf} \di{t_1} \int_0^{t_1} \di{t_2} \: \Bigl\{ \bigl[ \hV_{\uS,\uI}(t_1),
		\hW_{\uS,\uI}^{(1)}(t_2) \bigl] + \bigl[ \hW_{\uS,\uI}^{(1)}(t_1), \hV_{\uS,\uI}(t_2) \bigr] + \bigl[
		\hW_{\uS,\uI}^{(1)}(t_1), \hW_{\uS,\uI}^{(1)}(t_2) \bigr] \Bigr\}
	\end{aligned}
	\label{eq:alt1}
\end{equation}
In the general strategy presented in Sec.~\ref{sec:genth}, we neglect the term originating from the double integral with the
argument that it is a high order term in the perturbative series. However, if one calculates the commutators $[
\hH_{\mm{mod},\uI}^{(0)}(t_1), \hW_{\uS,\uI}^{(1)}(t_2) ]$ and $[ \hW_{\uS,\uI}^{(1)}(t_1), \hW_{\uS,\uI}^{(1)}(t_2) ]$ (cf.
Eqs.~\eqref{eq:VIS} and \eqref{eq:WIS} of the main text), one finds terms proportional to $\s{z} \ketbra{n}{n}$. This gives us a
path forward to find a correction Hamiltonian:~we will keep the higher order term originating from the double integral in
Eq.~\eqref{eq:alt1} and try to find a correction Hamiltonian $\hW_{\uS,\uI}^{(1)}(t)$ that guarantees that $\hOmega_1^{(1)}(t_\uf)
+ \hOmega_2^{(1)}(t_\uf) = 0$. 

Substituting the expression for the correction Hamiltonian in the interaction picture (see Eq.~\eqref{eq:WIS} of the main text) in
Eq.~\eqref{eq:alt1} and expanding the envelope functions  $g_{x,n}(t)$ and $g_{y,n}(t)$ in a Fourier series that we truncate at
$k=k_{max}$, we get a quadratic system of equations in the free parameters that allows us to satisfy the condition
$\hOmega_1^{(1)}(t_\uf) + \hOmega_2^{(1)}(t_\uf) = 0$.

Solving such a system of equations is still a difficult thing to do, since we have a system of $3 N_\mm{trunc}$ quadratic
equations depending on $4 k_\mm{max} N_\mm{trunc}$ free parameters. Here, $N_\mm{trunc}$ is the dimensionality of the truncated
cavity Hilbert space. There is, however, a convenient approximation one can do to simplify the problem:~one can assume that the
effect of $g_{x,n}(t)$ and $g_{y,n}(t)$ on cavity levels other than $\ket{n}$ is small and can be neglected. This allow us to
break the initial system of $3 N_\mm{trunc}$ equations in $N_\mm{trunc}$ independent systems of $3$ equations each, depending on
$4 k_\mm{max}$ free parameters only. These systems of equations have, however, several solutions since they are nonlinear. To
choose the ``best'' solution, it is convenient to work with more free variables than equations and use Lagrange multipliers to
find solutions that minimize the norm of the vector of free parameters. Such systems
can easily be solved with numerical methods. In this work we have solved the system of quadratic equations using the package
HomotopyContinuation.jl~\cite{Sbreiding2018} available for the Julia programming language~\cite{Sbezanson2017}.

Note that even in the case where solutions for Eq.~\eqref{eq:alt1} exist, it is not guaranteed that we will be able to mitigate
the effects of the unwanted Hamiltonian $\hV_{\uS}(t)$. This is the case when the envelopes of the correction Hamiltonian are not
``small''. In such a case, even though the first two terms of the Magnus expansion are zero at $t = t_\uf$, higher order terms
become larger, spoiling the dynamics. In fact, one could ask what a ``small'' correction Hamiltonian means in this context, and we
do not have a straightforward answer to that. In the example that we showed in Fig.~\ref{fig:fig05} of the main text, the spectral
weight of the corrected pulse was about $1/3$ of the original pulse. This is not what usually one calls ``small'' for a perturbative
series, but this solution does mitigate the effects of the error-Hamiltonian.

One can use the above strategy to also correct higher order errors. By including higher order terms in Eq.~\eqref{eq:alt1}, one
gets a system of higher order polynomials to solve, with substantially more free parameters. It is, however, cumbersome to handle
these higher order terms. A more convenient approach is to assume that the correction Hamiltonian is small enough such that its effect on higher
order terms in the Magnus expansion can be neglected. If this assumption is true, we can attack this problem in the following
way:~assume that we have found a correction Hamiltonian $\sum_n \hW_\uI^{(n-1)}(t)$, that corrects errors up to order
$\epsilon_{\uS}^{n-1} = (\chi t_\uf)^{n-1}$. Thus,
\begin{equation}
	\sum\limits_{k=1}^{\infty} \Omega_k^{(n-1)}(t_\uf) = 0 + \mathcal{O}(\epsilon_{\uS}^n).
	\label{eq:magnusn-1}	
\end{equation}
We want to add a term $\hW_\uI^{(n)}(t)$ to the correction Hamiltonian such that
\begin{equation}
	\sum\limits_{k=1}^{\infty} \Omega_k^{(n)}(t_\uf) = 0 + \mathcal{O}(\epsilon_{\uS}^{n+1}).
	\label{eq:magnusn}
\end{equation}
We can write $\Omega_k^{(n)}(t_\uf)$ in terms of $\Omega_k^{(n-1)}(t_\uf)$ and $\hW_\uI^{(n)}(t)$. If we assume that the effect of
$\hW_\uI^{(n)}(t)$ on $\Omega_k^{(n)}(t_\uf)$ for $k > 2$ is negligible, then 
\begin{equation}
	\Omega_k^{(n)}(t_\uf) \approx \Omega_k^{(n-1)}(t_\uf) \text{ for } k > 2.
	\label{eq:Omega_approx}
\end{equation}
Substituting Eqs.~\eqref{eq:magnusn-1} and \eqref{eq:Omega_approx} in Eq.~\eqref{eq:magnusn}, we obtain (up to order
$\epsilon_{\uS}^n$)
\begin{equation}
	\hOmega_1^{(n)}(t_\uf) + \hOmega_2^{(n)}(t_\uf) + \sum\limits_{j>2} \hOmega_j^{(n-1)}(t_\uf) = 0.
\end{equation}
The above equation can be rewritten as
\begin{equation}
	\begin{aligned}
		\sum\limits_j \hOmega_j^{(n-1)}(t_\uf) &= i \int_0^{t_\uf} \di{t_1} \hat{W}_\uI^{(n)}(t_1) \\
		& + \frac{1}{2} \int_0^{t_\uf} \di{t_1} \int_0^{t_1} \di{t_2} \: \Bigl\{ \bigl[ \hH_{mod,\uI}^{(n-1)}(t_1),
		\hat{W}_\uI^{(n)}(t_2) \bigl] + \bigl[ \hat{W}_\uI^{(n)}(t_1), \hH_{mod,\uI}^{(n-1)}(t_2) \bigr] + \bigl[
		\hat{W}_\uI^{(n)}(t_1), \hat{W}_\uI^{(n)}(t_2) \bigr] \Bigr\}.
	\end{aligned}
	\label{eq:alt2} 
\end{equation}
The sum on the left hand side runs over the Magnus terms whose leading order is $\epsilon_{\uS}^n$. Here it is useful to simply
replace the left hand side sum by a sum running from $j=1$ to $j=2n$ (see Appendix \ref{app:NonresonantPerturbations}). We can
then find $\hat{W}_\uI^{(n)}(t)$ using the methods discussed previously. We used this method to find fourth order corrections for
the SNAP problem shown in the main text.

\subsection{Correction Hamiltonian when Adding Phases on All Bosonic Number States}
\label{app:SNAPcorrection2}

In the situation in which one wants to have arbitrary phases imprinted in all bosonic number states of the truncated bosonic
Hilbert space, one can use the general method described in Sec.~\ref{sec:genth}. 

Since we want to have phases on all bosonic number states of the truncated bosonic Hilbert space, we need to have driving
components resonant with all the frequencies $\omega_{\uq,n} = \omega_\uq + n \chi$, for $n = 0, 1, \ldots, N-1$. Therefore, the
driving Hamiltonian in the lab frame is given by Eq.~\eqref{eq:HSNAPd2} of the main text. Thus, in the interaction picture with
respect to $\hH_{\uq\uc}$ (see Eq.~\eqref{eq:Hqc} of the main text), the Hamiltonian describing the dispersive coupling between
the qubit and cavity is given by $\hH_\uS(t) = \hH_{\uS,0}(t) + \hV_\uS(t)$, where
\begin{equation}
	\hH_{\uS,0}(t) = \sum\limits_{n=0}^{N-1}  \bigl[ f_{x, n}(t) \s{x} - f_{y,n}(t) \s{y} \bigr] \ketbra{n}{n},
	\label{eq:H0snap2}
\end{equation}
and
\begin{equation}
	\begin{aligned}
		\hV_\uS(t) = \textstyle \sum\limits_{n=0}^{N-1} & \textstyle \sum\limits_{n'=0}^{N-1} \Bigl( f_{x,n}(t) \bigl\{
		\cos[\chi (n' - n) t] \s{x} - \sin[\chi (n' - n_0) t] \s{y} \bigr\} \\
		& \qquad -f_{y,n_0}(t) \bigl\{ \sin[\chi (n' - n) t] \s{x} + \cos[\chi (n' - n) t] \s{y} \bigr\} \Bigr) (1 -
		\delta_{n',n}) \ketbra{n'}{n'},
	\end{aligned}
\end{equation}

Considering the same correction Hamiltonian as in the main text [see Eqs.~\eqref{eq:Wlabsnap} and \eqref{eq:WS}], we find that in
the interaction picture with respect to $\hH_{\uS,0}(t)$ [see Eq.~\eqref{eq:H0snap2}] $\hW_{\uS,\uI}(t)$ has terms proportional to
$\s{z} \ketbra{n}{n}$. In contrast to the case we considered in the main text, where $\hH_{\uS,0}(t)$ acts only on the subspace of
$\ket{n_0}$ (see Eq.~\eqref{eq:H0snap} of the main text), when we drive resonantly all bosonic number states the Hamiltonian
$\hH_{\uS,0}(t)$ [see Eq.~\eqref{eq:H0snap2}] acts on the whole (truncated) Hilbert space. This allow us to find a correction
Hamiltonian by simply solving a system of linear equations, like we did for the other problems discussed in the Applications
sections of the main text.

\end{appendix}

\end{document}